\documentclass[showpacs,twocolumn,aps,superscriptaddress,preprintnumbers,letterpaper]{revtex4-1}
\usepackage{amsmath,amssymb}
\usepackage{epsfig}
\usepackage{graphicx}
\usepackage{amsmath}
\usepackage{slashed}
\usepackage{amsfonts}
\usepackage{epstopdf}
\usepackage{color}
\usepackage{caption}
\usepackage{subcaption}
\usepackage[pdftex, pdfborder= 0 0 0, citecolor=blue, urlcolor=blue, linkcolor=blue, colorlinks=true, bookmarksopen=true]{hyperref}

\newcommand{\be}{\begin{equation}}
\newcommand{\ee}{\end{equation}}
\newcommand{\bear}{\begin{eqnarray}}
\newcommand{\eear}{\end{eqnarray}}
\newcommand{\ba}{\begin{array}}
\newcommand{\ea}{\end{array}}

\def\be{\begin{eqnarray}}
\def\ee{\end{eqnarray}}
\def\bea{\be}
\def\eea{\ee}

\def\roughly#1{\mathrel{\raise.3ex\hbox{$#1$\kern-.75em%
\lower1ex\hbox{$\sim$}}}}

\begin{document}

\title{Pion and Kaon parton  distributions in the QCD instanton vacuum}

\author{Arthur Kock}
\email{kock.arthur@stonybrook.edu}
\affiliation{Department of Physics and Astronomy, Stony Brook University, Stony Brook, New York 11794-3800 USA}

\author{Yizhuang Liu}
\email{yizhuang.liu@sju.edu.cn}
\affiliation{$^1$Tsung-Dao Lee Institute, Shanghai Jiao University, Shanghai, 200240, China}

\author{Ismail Zahed}
\email{ismail.zahed@stonybrook.edu}
\affiliation{Department of Physics and Astronomy, Stony Brook University, Stony Brook, New York 11794-3800, USA}



\date{\today}
\begin{abstract}
We discuss  a general diagrammatic description of n-point functions in the QCD  instanton vacuum  that
resums planar diagrams, enforces gauge invariance  and spontaneously broken chiral symmetry.
We  use these diagrammatic rules to derive  the pion and kaon quasi-parton amplitude and distribution
functions at leading order in  the instanton packing fraction for large but finite momentum. The instanton and anti-instanton zero modes and
non-zero modes are found to contribute to the quasi-distributions, but the latter are shown to drop out
in the large  momentum limit. The pertinent pion and kaon  parton distribution amplitudes and functions  are
made explicit at the low renormalization scale fixed  by the inverse instanton size.
Assuming that factorization holds, the pion parton distributions are evolved to higher renormalization scales with one-loop DGLAP and compared to existing data.

\end{abstract}


\maketitle

\setcounter{footnote}{0}


\section{Introduction}

Light cone distribution amplitudes are central to the description of hard exclusive processes
with large momentum transfer. They account for the non-perturbative quark and gluon content
of a hadron in the infinite momentum frame. Using factorization, hard cross sections can be
split into soft partonic distributions convoluted with perturbativly calculable processes. The
soft partonic distributions are inherently non-perturbative. They can be extracted through moments
 using experiments~\cite{FF}, or more recently lattice simulations~\cite{JI, JI1}.


Several QCD lattice simulations have suggested that the bulk characteristics and correlations of the light
hadronic operators are mostly unaffected by lattice cooling~\cite{CHU} , strongly suggesting that semi-classical gauge
and fermionic fields maybe dominant in the ground state. At weak coupling, instantons and anti-instantons
are exact classical gauge tunneling configurations with large actions and finite topological charge which
support exact quark zero modes with specific chirality.

Extensive calculations both analytically and numerically~\cite{DP,IQCD,MF},
describing the QCD ground state as an ensemble of instantons and anti-instantons with hopping quark zero modes
where found to reproduce most of the cooled lattice simulations. In the quenched approximation, these calculations
can be organized by observing that the ensemble is characterized by a small packing fraction in the large $N_c$ limit
which is dominated by planar graphs.


The twist-2 pion distribution amplitude and function have been discussed in the context of
the QCD sum rules~\cite{M1},  bottom-up holographic models~\cite{BROD},  bound state re-summations~\cite{M3}, basis light front quantization
using an effective Nambu-Jona-Lasinio (NJL) interaction~\cite{BASIS}, covariant  NJL models with effective interactions~\cite{BRO},
 and the instanton model using non-local effective interactions and  modified dipole-like or gaussian quark masses time-like~\cite{LAW1,LAW2}.

The instanton model for the QCD vacuum is inherently space-like. It is amenable to QCD through semi-classics and allows for an organizational principle
that enforces  chiral ang gauge Ward identities.  It is well suited for the description of the bulk of the QCD vacuum
with its its flavor singlet axial and scale anomalies, and its mesonic and baryonic excitations through pertinent
Euclidean correlators~\cite{DP,IQCD,MF}.  However, its inherent Euclidean character makes it difficult to
characterize the non-perturbative time-like structure of its partonic  constituents as probed by deep inelastic scattering.

The recent suggestion put forth by Ji~\cite{JI} and its matching protocol~\cite{JI2}, to extract the light cone distribution functions from
equal-time quasi-distributions in Euclidean space, has been carried on the lattice with some reasonable  success~\cite{LATTICE}. This
formulation makes the instanton calculus amenable to a first principle
semi-classical analysis of the quasi-distributions, and therefore the light cone distributions by matching.
 Since the distribution functions obey 
sum  rules,  the  enforcement of the Ward identities is important. This can be sought
through a diagrammatic expansion and
power counting   around the planar approximation, both of which preserve chiral and gauge invariance.
 We note that recently, the quasi-pion distribution functions were analyzed
using some of the models described above~\cite{RAD,QUASI}.

The purpose of this paper is to revisit  the n-point  functions in the random instanton
vacuum (RIV)  in the planar approximation as in~\cite{POB}.  The latter resums a large class of instanton contributions
in the form of non-perturbative integral equations with full chiral and gauge symmetry.
 While these equations are in general involved and require a numerical
analysis, we will analyze them relying on the diluteness of the instantons and anti-instantons
in the QCD vacuum, where the packing fraction is small with  $\kappa\approx 10^{-3}$.
 All  calculations will be carried at leading order (LO)
and/or  next to leading  order (NLO) in $\alpha\approx \sqrt{\kappa}$.


This expansion  provides
an organizational principle and addresses some of the  shortcomings in~\cite{DP,MF} by
enforcing the axial and vector Ward identities in the planar approximation.  It turns out that the zero modes
and non-zero modes  are equally important in this enforcement, as initially noted for
the 2-point vector correlator in~\cite{GROSS}. Also, the virtual character of the induced effective quark constituents
generated by the planar resummation makes their time-like manifestation in the pion distribution amplitude and function
very subtle.


This paper consists of several new results in the QCD random instanton vacuum (RIV):
1/ a generalization of the  planar resummation to the n-point functions;
2/ an explicit derivation of the 2- and 3-point functions at NLO;
3/ a derivation of the quark effective mass at NLO;
4/ a planar re-summation of the pion quasi-parton amplitude and distribution functions at LO;
5/ explicit expressions for the pion and kaon PDA, PDF and TMD  at LO;
6/ an explicit  proof of the axial Ward identity for the axial-axial correlation function at NLO;
7/ an explicit evolution of the pion PDA, PDF and TMD and comparison with currently available data.

The organization of the paper is as follows: in section II we briefly review the general aspects of  the random
instanton vacuum,  and detail the planar approximation  for the derivation of the quark propagator. We derive
the induced effective quark mass and in general spin-valued self-energy at NLO. In section III we show how
the 2-point meson correlators are re-summed, and use the result to derive the pion pseudo-scalar vertex
at NLO. The pion decay constant is worked out in the leading logarithm approximation. In section IV we
derive the pion QPDA  in LO  in the leading logarith approximation and beyond.  In section V we show
how the planar re-summation applies to the 3-point functions and use it to construct the pion QPDF, PDF and TMD also at LO.
All the results are QCD evolved and compared to available data.
Our conclusions are in section VI. A number of details can be found in several Appendices, including the LO result
for the GPDF.

\begin{widetext}

\section{Quark propagator}

Key to the analysis of the spontaneous breaking of chiral symmetry in the random instanton  vacuum  is the
appearance of a momentum dependent constituent mass dynamically. To illustrate this, consider the quark
propagator in the chiral limit and the quenched approximation of QCD

\be
\label{1}
\left<\psi(x)\psi^\dagger (y)\right>=\bigg<\left<x|(-i\slashed{\partial}-\slashed{A}-im)^{-1}|y\right>\bigg>_A
\ee
where the averaging is over the gauge configurations $A$. In the instanton vacuum, the gauge configurations
are restricted to the semi-classical set of instantons and anti-instantons which are sampled either random
(random instanton model) or through their semi-classical interactions (interacting instanton model). Throughout,
we will refer to the instanton liquid model as the random instanton model.  With this in mind and to enforce topological neutrality of the
vacuum, the semi-classical configurations are chosen to be $\frac N2$ instantons and $\frac N2$
anti-instantons non-interacting and randomly distributed in a 4-volume $V_4$, with

\be
\label{2}
A=\sum_{I=1}^{\frac N2}A_I+\sum_{\bar I=1}^{\frac N2}A_{\bar I}
\ee
Each of the instanton and anti-instanton $A_{I, \bar I}$ in (\ref{2}) is an SU(2) valued matrix embedded in
SU($N_c$) with arbitrary orientation $U_{I, \bar I}$ in color space and position $z_{I, \bar I}$ in 4-space.
For simplicity, all instantons are chosen to carry the same size
$\rho\approx 1/3$ fm with a density fixed at $n=N/V_4\approx  2\,{\rm fm}^{-4}$. In Fig.~\ref{size} we show the
instanton size distribution versus their size $\rho$ in fm from the instanton liquid model in comparison to
SU(2) lattice simulations~\cite{IQCD}\cite{MICHAEL}.

\begin{figure}[!htb]
 \includegraphics[height=6cm]{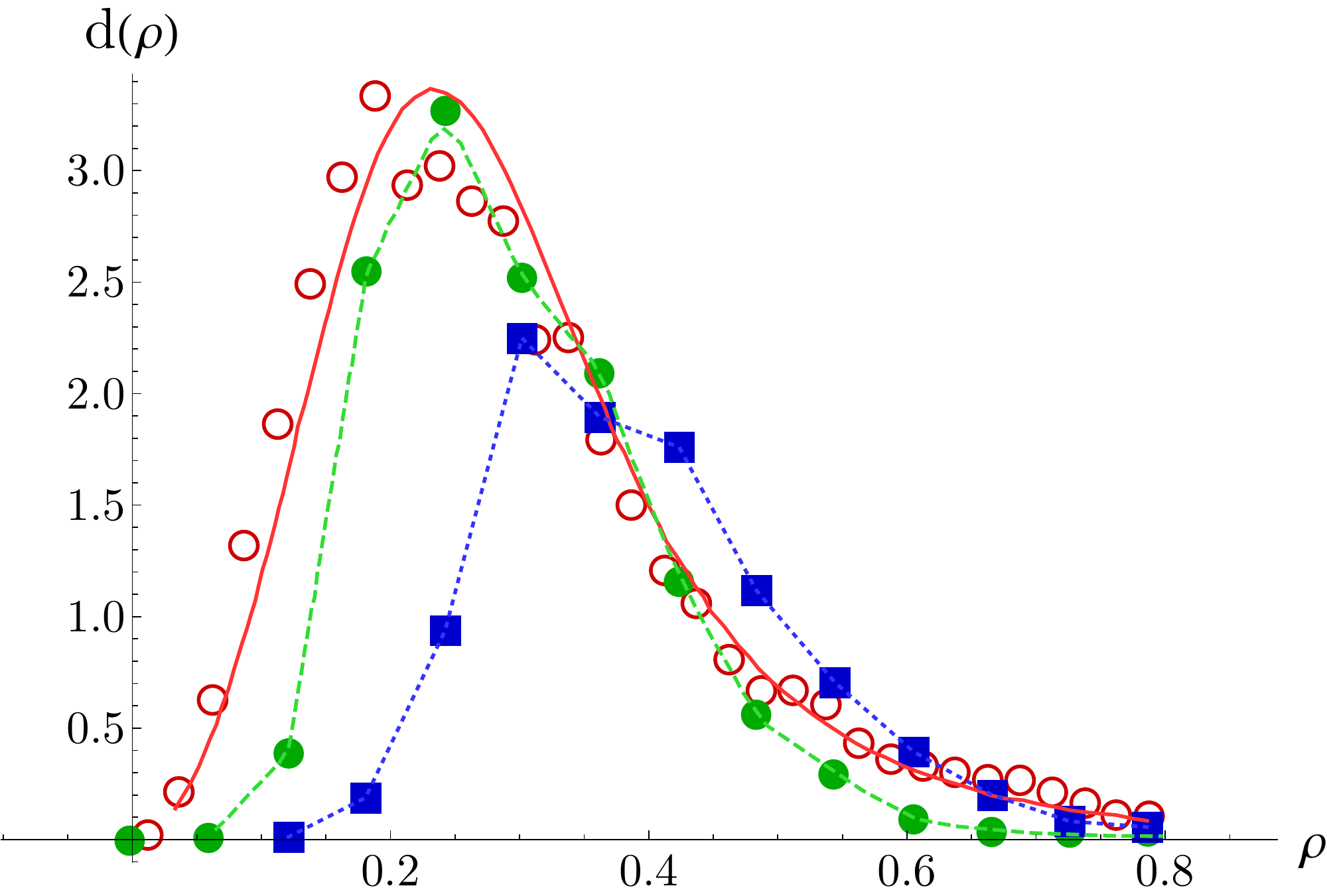}
   \caption{Instanton size distribution in pure SU(2) gauge theory calculated on the lattice \cite{MICHAEL} and in the interacting instanton liquid model (IILM) \cite{IQCD}. The unit scale is fm. Closed points correspond to lattice data with decreasing lattice spacing. Open points are from an IILM calculation. The solid line is a parameterization of the IILM result as discussed in \cite{IQCD}. The dotted and dashed lines simply aid the eye. In both pure SU(2) and pure SU(3), an average value $\rho\approx 1/3\, \textrm{fm}$ is found.}
    \label{size}
\end{figure}


In the semiclassical background (\ref{2}) the quark propagator can be organized as an expansion in
multiple re-scatterings with increasing number of instantons and anti-instantons

\be
\label{2X}
S-S_0=
\bigg<\sum_{ I}S_I-S_0
+\sum_{I\ne J}(S_I-S_0)S_0^{-1}(S_J-S_0)
+\sum _{ I\ne J, J\ne K}(S_I-S_0)S_0^{-1}(S_J-S_0) S_0^{-1}(S_K-S_0)
+..\bigg>
\ee
Here $S_0=1/(-i\slashed{\partial}-im)$ is the free quark propagator, $S_I=1/(-i\slashed{\partial}-\slashed{A}_I-im)$ is the
quark propagator in a single instanton or anti-instanton background  and the sum
is over {\bf all} instantons and anti-instantons. The averaging means independent
integrations over  the instanton and anti-instanton positions and color orientations.

\subsection{Planar re-summation}


Large $N_c$ QCD is essentially a quenched approximation  dominated by planar graphs. The
same applies to its semi-classical approximation in terms of a random instanton vacuum, if we note that $n\sim N_c$ and that the averaging
over the SU($N_c$) color orientations bring extra factors of $1/N_c$.  The planar contributions to the
quark propagator are re-summed  through the formal equation~\cite{POB}

\be
\label{2X1}
S-S_0=\frac{N}{2V_4}\int_{I+\bar I}dz_IdU_I\left(\frac 1{1-S_{\slashed{I}}\Delta}\right)S_{\slashed{I}}S_0^{-1}S
=\frac{N}{2V_4}\int_{I+\bar I}dz_IdU_I\,S_0\left(\frac 1{S_0S_{\slashed{I}}^{-1}S_0-S_0\Delta S_0}\right)S
\ee
with the single instanton (anti-instanton) propagator $S_I=S_0+S_{\slashed{I}}$ , and   the amputated
and re-summed propagator

\be
S_0^{-1}SS_0^{-1}=\Delta+S_0^{-1}
\ee
The integrations in (\ref{2X1})  is over the instanton moduli for fixed
instanton size. The integration over the rigid gauge color $U_I$ projects onto the color singlet channel,
while the integration over the global 4-position $z_I$ restores translational invariance.
(\ref{2X1}) is readily recast in the form

\be
\label{2X2}
S^{-1}-S_0^{-1}=\frac{N}{2V_4} \int_{I+\bar I}dz_IdU_I
\left(\frac 1{S-S_0(S_0^{-1}+S_{\slashed{I}}^{-1})S_0}\right)
\ee
Inserting  the identity

\be
S_{\slashed{I}}^{-1}=\frac 1{S_I-S_0}=S_0^{-1}(-A_I)^{-1}S_I^{-1}
\ee
in (\ref{2X2}) yields  a formal integral equation for the self-energy~\cite{POB}

\be
\label{3}
S^{-1}-S_{0}^{-1}=\frac{N}{2V_4}\int_{I+\bar I} dz_I dU_I\frac{1}{S-\slashed{A_I}^{-1}}\equiv \frac{N}{2V_4}\int_{I+\bar I} dz_I dU_I\,\Sigma_I
\ee
which sums over a {single} instanton plus anti-instanton. The SU($N_c$) averaging
in (\ref{3}) over $U_I$ projects onto the color singlet channel, and the z-integral restores translational invariance.
Throughout and for simplicity, only the massless case will be considered with $m=0$ unless specified otherwise.

For $S^{-1}-S_0^{-1}=-i\sigma$, (\ref{3}) takes the formal gap-like form

\be
\label{3X}
i\sigma = \frac{N}{2N_cV_4}\int_{I+\bar I} dz_I\,{\rm Tr}_C\bigg(\slashed{A_I} \frac{1}{i\slashed{\partial}+\slashed{A_I}+i\sigma}(i\slashed{\partial}+i\sigma)\bigg)
\ee
with ${\rm Tr}_C$ referring to the color trace (in some places below it will also mean spin as well).
The small parameter $\kappa$ or packing fraction,

\be
\label{4}
\kappa=\frac {N\rho^4}{2V_4N_c}\equiv \alpha^2 \rho^4\approx 3.186\,\times 10^{-3}
\ee
 is of order $N_c^0$ provided that the instanton density is made to scale as $N/V_4\sim N_c$.


\begin{figure}[!htb]
\includegraphics[width=12cm]{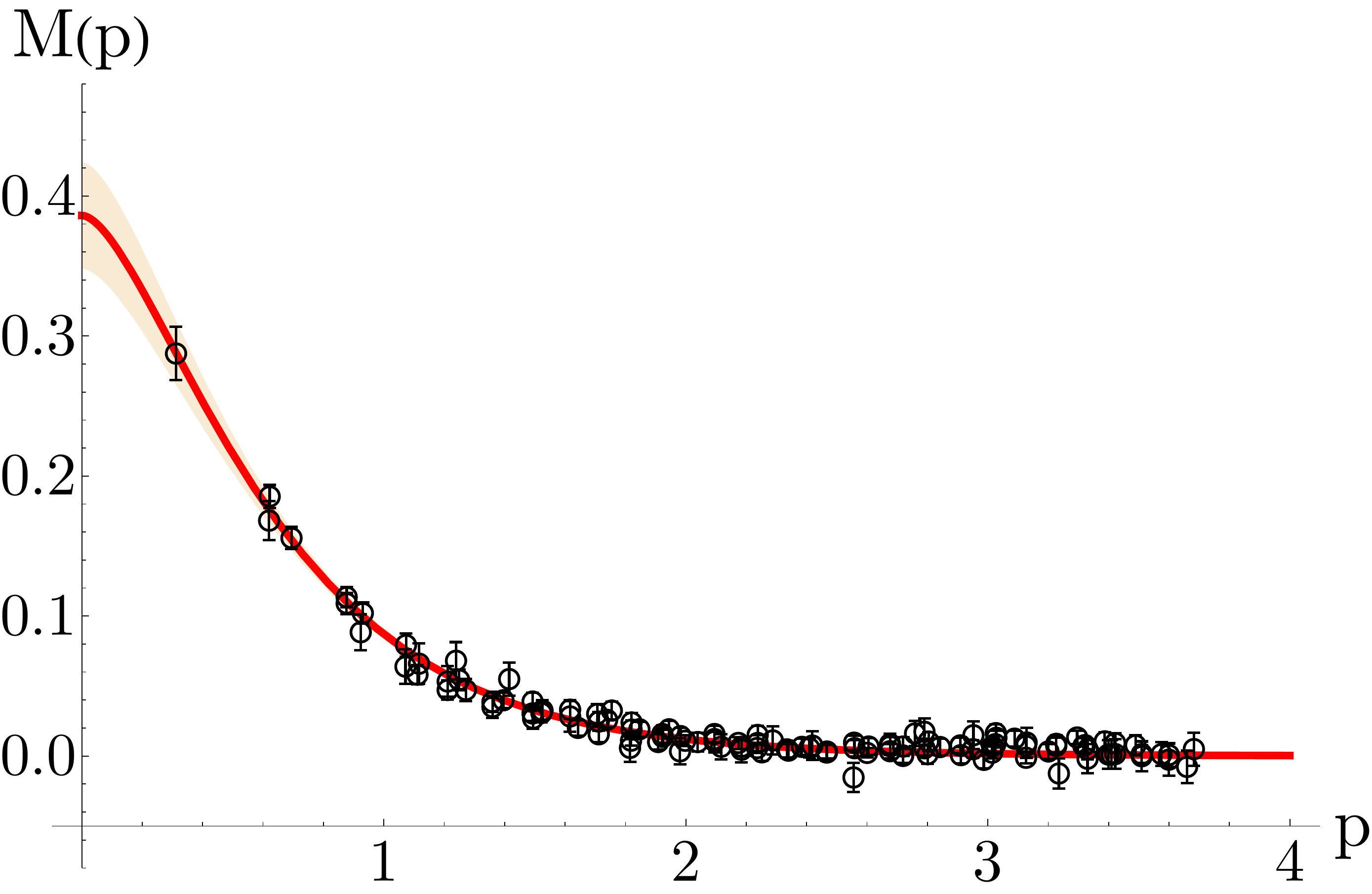}
  \caption{Momentum dependence of the 
  instanton induced effective  quark mass in singular gauge (\ref{MASSZ}) at LO  (solid-curves), compared to the effective quark mass 
  measured on the lattice in Coulomb gauge~\cite{BOWMAN} (open-circles). The unit scale is GeV. We obtain a fitted parameter intervals $M(0)=383\pm 39$ MeV and $\rho = 0.313 \pm 0.016$ fm.}
  \label{mass}
\end{figure}

\subsection{Effective mass at LO}

The effective mass operator (\ref{3}) can be sought iteratively $\sigma=\alpha\sigma_0+\alpha^2\sigma_1+...$, with
a starting correction of order $\alpha$ and not $\alpha^2$~\cite{POB}.  In leading order (LO),

\be
\label{4X}
\Delta_I(x,y)\equiv \left<x\right| \frac{1}{i\slashed{\partial}+\slashed{A_I}+i\alpha \sigma_0 }\left|y\right>=
\frac{\psi_{0I}(x)\psi_{0I}^\dagger(y)}{i\alpha\sigma_{00}}+{\cal O}(\alpha)
\ee
with $\sigma_{00}=\left<0|\sigma_0|0\right> =\sqrt{2}||q\varphi^{\prime 2}||\sim \rho$,
 the expectation value of $\sigma_0$ in the zero mode state $\psi_{0I}$. 
Its form is given in Appendix A both in regular and singular gauge. The norm notation is subsumed.
 Inserting (\ref{4X}) into (\ref{3X}) yields  the effective mass at LO

\be
\label{4X1}
M(p)\approx \alpha\sigma_0(p)=\frac \alpha {\sigma_{00}}{\rm Tr}_C
\left(\slashed{p}\left(\psi_{0I}(p)\psi_{0I}^{\dagger}(p)+\psi_{0\bar I}(p)\psi_{0\bar I}^{\dagger}(p)\right)\slashed{p}\right)
\equiv
\frac \alpha {\sigma_{00}}{\rm Tr}_{C}
\left({\slashed{p}\left( P_I(p)+P_{\bar I}(p)\right)\slashed{p}}\right)=\frac \alpha{\sqrt 2}\frac{|p\varphi^{\prime }(p)|^2}{||q\varphi^{\prime 2}||}\nonumber\\
\ee
with $P_I$ the zero mode projector.
 After color tracing, the effective mass operator is diagonal in spin.
The last relation holds for both the regular and singular gauge, but with different zero mode profiles.
In singular gauge from Appendix A we have 

\be
\label{MASSZ}
\frac{M(p)}{M(0)}=\bigg(\left|z\left(I_0K_0-I_1K_1\right)^\prime\right|^2\bigg)_{z=\frac 12 \rho p}
\ee
The effective mass is not analytic in the packing fraction since  $\alpha\sim \sqrt{\kappa}$.
In Fig.~\ref{mass}  we show (\ref{4X1}) for $M(0)=383$ MeV and $\rho=0.313$ fm in solid-red curve
in comparison to the  lattice generated effective quark mass in Coulomb gauge~\cite{BOWMAN}.
The spread corresponds to a $99\%$ confidence interval generated by a standard weighted least squares regression on the data in Fig.~\ref{mass}, giving parameter ranges $M(0)=383\pm 39$ MeV and $\rho = 0.313 \pm 0.016$ fm. 
At   large momenta,  $M(p)\approx 1/p^{6}$ falls faster than $\alpha^D_s(p^2)/p^2$ with $D=39/27$  derived using  short distance
QCD arguments~\cite{POLITZER}, but in good agreement with the lattice results.

\subsection{Effective mass at NLO}

At next to leading order (NLO) the effective mass is
obtained by further expanding (\ref{4X}) and keeping  the contributions to order
$\alpha$. For that, we can organize (\ref{4X}) formally so that

\be
\label{NLO1}
\Delta_I=&& \frac{1}{i\slashed{\partial}+\slashed{A_I}+i\alpha \sigma_{00} } +
\frac{1}{i\slashed{\partial}+\slashed{A_I}+i\alpha \sigma_{00} }(-i\alpha (\sigma_0-\sigma_{00}))\,\Delta_I\nonumber\\
=&& \bigg(\frac{P_I}{i\alpha\sigma_{00}} +G_I\bigg)+ \bigg(\frac{P_I}{i\alpha\sigma_{00}} +G_I\bigg)
(-i\alpha (\sigma_0-\sigma_{00}))\,\Delta_I
\ee
with $P_I$ the projector on the quark zero mode, and $G_I$ the quark non-zero mode propagator in the instanton background. Using the virtual quark eigenstates
$(i\slashed{\partial}+\slashed{A_I})\psi_{kI}=E_k\psi_{kI}$ and the mode expansion for
$G_I=\sum_{k\ne 0} \psi_{kI}\psi^\dagger_{kI}/E_k$ we can explicitly solve for $\Delta_I$ at NLO

\be
\label{NLO2}
\Delta_I=&& \frac{P_I}{i\alpha\sigma_{00}}+
 \sum_{k\ne 0} \frac{\psi_{kI}\psi_{kI}^{\dagger}}{E_k}+P_I\sum_{k \ne 0}\frac{\sigma_{0k}\sigma_{k0}}{E_k\sigma_{00}}-\sum_{k\ne 0}\frac{\psi_{kI}\psi_{0I}^{\dagger}\sigma_{k0}}{E_k \sigma_{00}}-\sum_{k\ne 0}\frac{\psi_{0I}\psi_{kI}^{\dagger}\sigma_{0k}}{E_k \sigma_{00}} +{\cal O}(\alpha)
\ee
The matrix elements $\sigma_{kl}=\left<k|\sigma|l\right>$ involve $\sigma$ only in leading order or  $\sigma_0$. From (\ref{3X}) the
self-energy at NLO is spin-valued and reads

\be
\label{NLO3}
\Sigma_I&=&\slashed{A_I} \frac{-1}{i\slashed{\partial}+\slashed{A_I}+i\sigma}(i\slashed{\partial}+i\sigma)\nonumber\\
&\approx& i\slashed{\partial}\frac{P^I}{i\alpha\sigma_{00}}i\slashed{\partial}
+i\slashed{\partial}(1-P^I\hat \sigma )G^I(1-\hat \sigma P^I)i\slashed{\partial}-i\slashed{\partial}
+i\slashed{\partial}P^I \hat \sigma+\hat \sigma P^Ii\slashed{\partial}-\frac{\beta_{00}}{i\sigma_{00}^2}i\slashed{\partial}P^Ii\slashed{\partial}
\approx \frac{\Sigma_{I0}}{\alpha}+\Sigma_{I1}
\ee
with $\hat \sigma ={\sigma_0 }/{\sigma_{00}}$. The last  contribution was added to account for the planar re-summation  to the
self-energy through the gap-equation

\be
\label{NLO4}
-i\beta_{00}(p)=\int d^4x\,d^4y\,e^{ip\cdot (x-y)}\,{\rm Tr}_C\Sigma_{I1}(x,y)={\rm Tr}_C\Sigma_{I1}(p,p)
\ee
\end{widetext}
The  anti-instanton contribution follows through the substitution $I\rightarrow \bar I$. The
 effective mass at NLO follows by substituting (\ref{NLO3}) in (\ref{3X}) which is now a true gap-equation because of the
 contribution (\ref{NLO4}). The first contribution is in agreement with (\ref{4X1}).
 The non-zero mode contribution $G_I$ in (\ref{NLO3}) is important for  the enforcement of symmetries at NLO.
  Its explicit form in singular gauge  is given in Appendix A.

\section{Mesonic vertices in the planar approximation}

In this section we formulate the planar resummation for the 2-point functions.
We will focus on flavor singlet correlators with a single flavor and ignore vacuum loops,
with the assumption that we are evaluating non-singlet flavor correlators where the loop corrections
are suppressed by $1/N_c$. We will analyze in details the resummation for the 2-point pseudoscalar
source at LO and NLO.


\subsection{Meson correlators}

The T-matrix in the planar approximation for re-scattering of two quarks in the $2\rightarrow 2$
channel, can be constructed using formally the 2PI kernel

\be
\label{7}
K=\frac{N}{2V}\int_{I+\bar I} dz_I dU_I\,\bigg(\Sigma_I\otimes \Sigma_{I}\bigg)
\ee
with the color-spin-space valued self-energy $\Sigma_I$ given in (\ref{3}).
We can now use the 2PI kernel to construct any  2-point correlation function in the QCD instanton vacuum.
For that, consider the  general local and colorless source  $J_{\Gamma}=\psi^{\dagger}\Gamma\psi$, say for
a meson of spin-flavor $\Gamma$, and define the
amputated spin-flavored valued operator

\be
\label{13}
&&O_{\Gamma\, ab}(P,k)=\nonumber\\
&&S^{-1}(k)\left<\psi_a(k) \psi_b^\dagger(P-k) J_{\Gamma}(P)\right>S^{-1}(k-P)\nonumber\\
\ee
where the averaging is carried over the instanton-anti-instanton vacuum gauge configurations.
(\ref{13}) refers to a quark of color-spin-a and momentum $k$ combining with a quark of color-spin-b and
momentum $P-k$ to give a colorless meson of spin-$\Gamma$ and momentum $P$.

The resummed instanton contributions to (\ref{13}) in the planar approximation can be obtained
by tracing (\ref{13}) with the 2PI kernel (\ref{7}) with the result

\begin{widetext}
\be
\label{14}
O_{\Gamma}(P,k)=\Gamma
+ \alpha ^2\rho^4\sum_{I+\bar I} \int \frac{d^4p}{(2\pi)^4} \,{\rm Tr}_C\bigg(\Sigma_{I}(k,p)S(p)
 O_{\Gamma}(P,p)S(p-P)\Sigma_I(p-P,k-P)\bigg)
\ee
\end{widetext}
The net effect of averaging over the instanton-anti-instanton ensemble is the
projection onto the color singlet channel and momentum conservation.
(\ref{14}0 is the analogue of the Bethe-Salpeter equation for the vertex functions.
Using (\ref{14}) the re-summed 2-point function for an arbitrary meson correlator
 in the planar approximations is

\begin{widetext}
\be
\label{15}
{\cal C}_{\Gamma_1\Gamma_2}(P)\equiv\left<O_{\Gamma_1}(-P)O_{\Gamma_2}(P)\right>=
- \int  \frac{d^4k}{(2\pi)^4} \,{\rm Tr}_C\bigg(\Gamma_1 S(k)O_{\Gamma_2}(P,k)S(k-P)\bigg)
\ee
\end{widetext}

\subsection{Pseudo-scalar pion vertex at LO}

The leading-order $\alpha$ contribution to the pion pseudoscalar correlator can be obtained
by setting $\Gamma=\gamma_5$, $\Sigma_I\sim\Sigma_0/\alpha$ and  $S\sim1/(\slashed{k}-i\alpha \sigma_0(0))$
in (\ref{14}-\ref{15}). As we will show below, many of our expressions will in fact be (logarithmically) divergent in $\alpha$. We will encounter integrals of the form
($n\geq2$)

\be
\label{KPSHIFT}
\int \frac{d^4 k}{(2\pi)^4}\frac{F(k)}{\left(k^2+\alpha^2 \sigma_0^2(0)\right)^{n}}
\ee 
where $F(k)$ is a smooth function that approaches a constant value as $k^2\rightarrow 0$ and rapidly drops off as $k^2\gg\rho^{-2}$. To extract the leading 
contribution in $\alpha$ from the integral of this kind, we may shift $k^2\rightarrow k^2-\alpha^2\sigma_0^2(0)$, and subsequently drop corrections to $F(k^2-\alpha^2\sigma_0^2(0))\approx F(k^2)$ which only contribute subleading divergent corrections (e.g. $\alpha \log(\alpha)$, $\alpha^2 \log(\alpha)$, etc.). As will prove useful later, it is totally equivalent to instead only shift $k_{\perp}^2 \rightarrow k_{\perp}^2-\alpha^2\sigma_0^2(0)$, which then leads to a shift in the lower bound of integration of only $k_{\perp}^2$.

\be
\label{KPSHIFT2}
k_{\perp}^2>\alpha^2\sigma_0^2(0)=M^2(0)
\ee 

With this in mind, a formal solution follows for the vertex operator $O_5(P,k)\approx \gamma^5F_5(P,k)$,
with the vertex function $F_5(P,k)$ diagonal in spin space. In singular gauge,
the latter formally satisfies the integral equation

\begin{widetext}
\be
\label{15X}
\gamma^5F_5(P,k)=\gamma^5+\frac{\varphi^\prime(k)\varphi^\prime(k-P) |k||k-P|}{\sigma_{00}^2}\,
\int  \frac{d^4p}{(2\pi)^4}F_5(P,p)
\left(-\psi_I^\dagger(p)\psi_I(p-P)\frac{1-\gamma_5}2+\psi_{\bar I}^\dagger(p)\psi_{\bar I}(p-P)\frac{1+\gamma_5}2\right)\nonumber\\
\ee
\end{widetext}

This is a Fredholm integral equation of the second-kind, and can be solved with a Liouville-Neumann series, which is found to be geometric. The solution then follows by summation with the result

\be
\label{16}
F_5(P,k)=
1+\lambda(P)\frac{\varphi^{\prime}(k)\varphi^{\prime}(k-P)|k||k-P|}{\sigma_{00}^2}
\ee
with $\lambda(P)$ satisfying

\be
\label{17}
\lambda(P)\bigg(1-\frac 2{\sigma_{00}^2}\,{\int \frac{d^4 k}{(2\pi)^4}\varphi^{\prime 2}(k)\varphi^{\prime 2}(k-P)(k^2-k\cdot P)}
\bigg)=1\nonumber\\
\ee
For small momentum we have $\lambda(P)\approx (\sqrt{N_c}/f_\pi)/P^2$, corresponding to the pion pole in the chiral limit.
We note that the  LO contribution (\ref{16})  amounts to the effective coupling  at the pion pole

\bea
\label{15X0}
\bigg(iO_5(P,k)\bigg)_{P^2\approx 0}
\approx\frac{\sqrt{N_c}}{f_\pi} \sqrt{M(k)}\left(\frac{i\gamma_5}{P^2}\right)\sqrt{M(k-P)}\nonumber\\
\eea
with the running mass of order $\alpha$ given in (\ref{4X1}). This  LO result is in agreement
with the effective vertex following from the partially resummed planar diagrams  in~\cite{DP,IQCD,MF}.
(\ref{16}) gives the full pion pseudoscalar vertex on- and off-mass-shell in the massless limit. The massive case will
be discussed below.

\subsection{Pion decay constant $f_\pi$ at LO}

The explicit values of $g_\pi, f_\pi$ follow from
the pseudo-scalar two-point correlation function
(\ref{15}) with the vertex function (\ref{16}), which is of order $\alpha^0$
to LO. More specifically,  the pion pole contribution reads

\be
 \label{15X}
{\cal C}_{\gamma_5;\gamma_5}(P)\approx- \int \frac{d^4k}{(2\pi)^4}
{\rm Tr}\left(\gamma_5F_5(P, k)\frac 1{\slashed k}\gamma_5 \frac 1{\slashed{k}-\slashed{P}}\right)
\approx  \frac{g_\pi^2}{P^2}\nonumber\\
\ee
$g_\pi$ defines the pseudoscalar pion-quark-quark coupling and in LO  is given by

\begin{widetext}
\be
 \label{15X1}
g_\pi^2=-N_c
\left(\int \frac{d^4k}{(2\pi)^4}\left(3\,\varphi^{\prime 4}+
7(k\rho)\, \varphi^{\prime 3}\varphi^{\prime\prime}
+{(k\rho)^2}\left(\varphi^{\prime 3}\varphi^{\prime\prime\prime}+\varphi^{\prime 2}\varphi^{\prime\prime 2}\right)\right)\right)^{-1}
\ee
\end{widetext}
The pion decay constant $f_\pi$
is related to $g_\pi$ in (\ref{15X1}) by   chiral reduction with

\be
\label{COND}
f_\pi g_\pi =-2i\left<\psi^\dagger \psi\right>\approx 2N_c\frac{M(0)}{(2\pi \rho)^2}
\ee
in leading order in the current quark mass.
The last identity is the LO contribution to the chiral condensate, and is the expected
Goldberger-Treiman relation for the effective quark coupling.
(\ref{COND})  is infrared finite and of order $\alpha$. Both $f_\pi, g_\pi$ are infrared sensitive
with

\be
\label{15X1X}
g_\pi^2\approx  N_c \bigg(\int\frac{d^4k}{(2\pi)^4}\frac{(2\pi\rho)^4}{k^4}\bigg)^{-1}
\ee
and similarly for $f_\pi$

\be
\label{FPI2}
f_\pi^2\approx 4N_c\int\frac{d^4k}{(2\pi)^4}\, \frac{M^2(0)}{k^4}
\ee
Note that $g_\pi$ in (\ref{15X1}) and therefore $f_\pi$ in (\ref{COND}) are UV finite with an approximate range
of $1/\rho$. 
The infrared sensitivity follows from the shift which led to (\ref{KPSHIFT2}). This shift will be understood throughout.

To logarithmic accuracy $g_\pi^2\sim  N_c/ {\rm ln}(1/\alpha)$ and
$f_\pi^2\sim N_c\alpha^2{\rm ln}(1/\alpha)$ with 

\be
\label{FPI3}
f_\pi^2\approx \frac{N_cM^2(0)}{2\pi^2}\,{\rm ln}\bigg(\frac C{\rho M(0)}\bigg)\\\nonumber
\ee
modulo a constant $C$ of order 1. $f_\pi^2$ captures the chiral conductivity in the QCD vacuum~\cite{RANDOM}.
Using the values of $M(0)\rho$ displayed in Fig.~\ref{mass} we have $C=0.849$ for $f_\pi=86\,\textrm{MeV}$ in the chiral limit, 
$C=0.897$ for $f_\pi=93\,\textrm{MeV}$ for massive pions, and $C=1.04453$ for $f_K=110\,\textrm{MeV}$ for massive kaons 
(see below).





\begin{widetext}
\subsection{Pseudo-scalar pion vertex at NLO}

The pseudo-scalar pion vertex can be sought at NLO using

\be
\label{PPNLO}
O_5(P,k)=\gamma^5\bigg(1+F_5(P,k)\bigg) +\alpha \overline{F}_5(P,k)+{\cal O}(\alpha^2)
\ee
in (\ref{14})  and the explicit form of the spin-valued self-energy (\ref{NLO3}) at NLO. The result is

\be
\label{19X}
&&\overline{F}_5(P,k)=K_{\pi} \overline{F}_5(P,k)\nonumber\\
&&+\rho^4 \sum_{I+\bar I}\int \frac{d^4p}{(2\pi)^4}
{\rm Tr}_C\bigg(\Sigma_{I1}(k,p) S_0(p)\gamma^5 {F}_5(P,p)S_0(p_-)\Sigma_{I0}(p_-,k_-) +
\Sigma_{I0}(k,p) S_0(p)\gamma^5 {F}_5(P,p)S_0(p_-)\Sigma_{I1}(p_-,k_-)\bigg)\nonumber\\
\ee
with $k_-,p_-=k-P, p-P$.  The reduced kernel $K_{\pi}$ involves only the zero modes and  satisfies

\be
\label{20X}
K_{\pi} O= \rho^4 \sum_{I,\bar I}\int \frac{d^4p}{(2\pi)^4} {\rm Tr}_C\bigg( \Sigma_{I0}(k,p) S_0(p)O S_0(p_-)\Sigma_{I0}(p_-,k_-)\bigg)\nonumber\\
\ee
\end{widetext}
(\ref{19X}) defines formally an integral-type equation for the spin-valued operator $\overline{F}_5(P,k)$.  The homogeneous part does not iterate due to a mismatch in chirality.
In the inhomogeneous contribution, we note that most of the contributions in $\Sigma_{I1}$ as given in (\ref{NLO3}) do not contribute due to a mismatch in chirality except for
$\delta G_I=G_I-S_0$, thus  the result

\begin{widetext}
\be
\label{Z39X}
\overline{F}_5(P,k)=&&
 \sum_{I+\bar I}\int  \frac{d^4 p}{(2\pi)^4}{\rm Tr}_C\left(\slashed{k}\delta G_{I}(k,p)\gamma^5 F_5(P,p)
 \psi_{0I}(p_-)\frac 1{i\sigma_{00}}
 \psi_{0I}^{\dagger}(k_-)\slashed{k}_-\right)\nonumber\\
+&&\sum_{I+\bar I}\int  \frac{d^4 p}{(2\pi)^4}
{\rm Tr}_C\left(\slashed{k}\psi_{0I}(k)\frac 1{i\sigma_{00}}\psi^{\dagger} _{0I}(p)
 \gamma^5 F_5(P,p)\delta G_I(p_-,k_-)\slashed{k}_-\right)
\ee
%
It may be checked that the non-vanishing spin-valued structures of  (\ref{Z39X}) are  of  the type $\gamma_5\slashed P$ and $\gamma_5\slashed k$
times invariant scalars.  For estimates, we may use  $\delta G_I$ in  the Born approximation (singular gauge)

\be
\label{BORN}
\delta G_I(k,p)\rightarrow \frac 1{S_0^{-1}+i\slashed A}_I-S_0\approx -iS_0\slashed A_I S_0\approx iS_0
\bigg(\frac{\gamma^M{\bar\sigma}^{MN}x^N\rho^2}{(x^2+\rho^2)x^2}\bigg)S_0
\rightarrow 4\pi^2\frac 1{\slashed{k}}\bigg(\gamma^M{\bar\sigma}^{MN}q^N\frac {(q\rho )^2K_2(q\rho )-2}{q^4}\bigg)\frac 1{\slashed{p}}\nonumber\\
\ee
\end{widetext}
with $q=|p-k|$.
We note that asymptotically $G_I\approx S_0$, which prompts the use of $\delta G_I\approx 0$ in most calculations
in the random instanton model. The Born approximation allows to go beyond.


\section{Pion quasi-parton distribution amplitude}

The most extensively studied partonic distribution is the twist-2 pion parton distribution amplitude
(PDA) which characterizes the  amplitude to find a pair of $q,\bar q$ with parton fraction $x,\bar x$ of the
pion total longitudinal momentum and $x+\bar x=1$.
The PDA  is constrained by the empirical pion form factor~\cite{FF}
and is known at asymptotic scales to be $6x\bar x$~\cite{AS}.  At lower scales,  there are model
calculations~\cite{BRO,LAW1,LAW2}. Recently, a QCD lattice simulations was used to extract the pion
quasi-parton distribution amplitude (QPDA) based on the large momentum effective theory~\cite{JI}
following the original  suggestion in~\cite{JI1}.

The proposed quasi-parton distribution put forth in~\cite{JI1}, translates to the pion QPDA
for the twist-2  as

\begin{widetext}
\be
\label{X1}
\tilde  \phi_\pi(x, P_z)=\frac i{f_\pi}\int \frac{dz}{2\pi}\,e^{-i(x-\bar x)zP_z/2}\,
 \left<\pi(p)\right|{\psi}^\dagger(z_-)\gamma^z\gamma^5\,[z_-,z_+]\,
\psi(z_+)\left|0\right>
\ee
\end{widetext}
where the quark fields are separated along the z-direction at $z_\pm=\pm z/2$
in Euclidean space,  and $[z_-,z_+]$ is a  gauge link enforcing gauge invariance.
Gauge links  in Euclidean space correspond to heavy quark propagators. In the single
instanton or  anti-instanton background they are defined in Appendix C. Long links
develop a  self-energy in the form
$e^{-\Delta z}$,  with generically $\Delta\sim \alpha\rho$
and typically $\Delta\approx 70$ MeV~\cite{SELF}.
Note that in the infinite momentum limit this contribution is of order
$e^{-\Delta/P_z}\approx 1$.

The amplitude (\ref{X1}) is normalized by the PCAC condition

\be
\label{X1X}
\int_{-\infty}^{+\infty} dx\,\tilde\phi_\pi(x, P_z)=\frac i{f_\pi P_z}
\left<\pi(p)|{\psi}^\dagger(0)\gamma^z\gamma^5\psi(0)|0\right> =1\nonumber\\
\label{X2}
\ee
The pion light cone distribution amplitude follows by taking the limit $P_z\rightarrow \infty$
(infinite momentum) through
perturbative matching~\cite{JI2}. We note that $x, \bar x$ are in general unbound with $0\leq x, \bar x\leq 1$
only expected in the infinite momentum limit.  More general properties of the QPDA were recently discussed in~\cite{RAD}.
A more general QPDA is discussed in Appendix D.

\subsection{Planar approximation}

In the random instanton vacuum, a typical  planar contribution to  the matrix element in (\ref{X1})
is illustrated in Fig.~\ref{pcac1}. It follows from the 2-point like correlator
$\left<J_5{\mathbb J}_{5z}\right>$ with ${\mathbb J}_{5z}$ a point split non-local source.
If we set the gauge link in (\ref{X1}) to 1  as we argued earlier, the properly normalized result at the pion pole is

\begin{widetext}
\be
\label{X3}
\tilde\phi_\pi(x,P_z)=\lim_{P^2\rightarrow 0}\frac {-i}{f_\pi g_\pi}\frac{P^2}{P_z}
\int \frac{d^4k}{(2\pi)^4} \delta \left(x-\frac{1}{2}-\frac {k_z}{P_z}\right){\rm Tr}_C\left(\gamma^z\gamma^5S_1
O_5(p_1,p_2)S_2\right)
\ee
where $p_{1,2}=k\pm \frac P2$  and

\be
\label{X4}
p_1^2=\left(k_4\pm \frac {i}2 E_\pi\right)^2+k_\perp^2+x^2P_z^2\\
p_2^2=\left(k_4\mp \frac {i}2 E_\pi\right)^2+k_\perp^2+\bar x^2P_z^2
\ee
with $E_\pi=P_z$.  We note that (\ref{X3}) is of order $\alpha^0$ since the trace-part is of order $\alpha$ and
$f_\pi\sim \alpha$ from (\ref{FPI2}). Specifically, using the pseudo-scalar vertex at NLO (\ref{PPNLO}), we have at the pion pole

 \be
\label{X4}
\tilde\phi_\pi(x,P_z)&&\approx\lim_{P^2\rightarrow 0}\frac {- i}{g_\pi f_\pi }\frac{P^2}{P_z}
\int \frac{d^4k}{(2\pi)^4} \delta \left(x-\frac{1}{2}-\frac {k_z}{P_z}\right)
{\rm Tr}_C\left(\gamma^z\gamma^5
\frac 1{{\slashed p}_1}\gamma^5F_5(p_1,p_2)\frac  {i\alpha\sigma_0({p_2})}{{p}^2_2}\right)\nonumber\\
&&+\lim_{P^2\rightarrow 0}\frac {-i}{g_\pi f_\pi }\frac{P^2}{P_z}
\int \frac{d^4k}{(2\pi)^4} \delta \left(x-\frac{1}{2}-\frac {k_z}{P_z}\right)
{\rm Tr}_C\left(\gamma^z\gamma^5
\frac {i\alpha\sigma_0({p_1})}{{p}^2_1}\gamma^5F_5(p_1,p_2)\frac  1{{\slashed p}_2}\right)\nonumber\\
&&+\lim_{P^2\rightarrow 0}\frac {-i}{g_\pi f_\pi }\frac{P^2}{P_z}\int \frac{d^4k}{(2\pi)^4} \delta \left(x-\frac{1}{2}-\frac {k_z}{P_z}\right)
{\rm Tr}_C\left(\gamma^z\gamma^5
\frac 1{{\slashed p}_1}\alpha\overline{F}_5(p_1,p_2)\frac 1{{\slashed p}_2}\right)
\ee
where the trace is now over color-spin.
Inserting  the pseudoscalar vertices at NLO (\ref{16}) and (\ref{Z39X}) in (\ref{X4}) give  the leading contribution of order $\alpha^0$ to the
QPDA

\be
\label{X7X}
\tilde\phi_{\pi}(x,P_z)\approx&&
-\frac {4 N_c }{f^2_\pi}\int \frac{d^4k}{(2\pi)^4}\delta \left(x-\frac{1}{2}-\frac {k_z}{P_z}\right)
\,\left({M({p_1})M({p_2})}\right)^{\frac 12}\, \left(\frac{\bar xM({p_1})+{x}M({p_2})}{p_1^2p_2^2}\right)
\nonumber\\
&&+\lim_{P^2\rightarrow 0}\frac{\alpha}{g_\pi f_\pi \sigma_{00}} \frac{P^2}{P_z}
\sum_{I+\bar I}\int  \frac{d^4 k d^4q}{(2\pi)^8}\delta \left(x-\frac{1}{2}-\frac {k_z}{P_z}\right) \nonumber \\
&&\times\bigg({\rm Tr}_C\bigg(\gamma^z\gamma^5 \delta G_I(p_1, q_1)\gamma^5F_5(q_1,q_2)\psi_{0I}(q_2)\psi_{0I}^{\dagger}(p_2)\bigg)\nonumber \\
&&+{\rm Tr}_C\bigg(\gamma^z \gamma^5 \psi_{0I}(p_1)\psi_{0I}^{\dagger}(q_1)\gamma^5 F_5(q_1,q_2)
\delta G_{I}(q_2, p_2)\bigg)\bigg)\nonumber\\
\ee
with $p_{1,2}=k\pm P/2$, $q_{1,2}=q\pm P/2$, and  $E_\pi=(P_z^2+m_\pi^2)^{\frac 12}\rightarrow P_z$ in the chiral  limit.
The first contribution involves only the zero modes, while the second contribution involves the cross contribution from
zero modes and non-zero modes.
With the help of the axial Ward identity, we have  checked that to order $\alpha^0$, (\ref{X7X}) with the link modification
(see below)  is properly normalized,

\be
\label{X8}
\int_{-\infty}^{+\infty} dx\,\tilde \phi_{\pi}(x,P_z)=1
\ee




\begin{figure}[!htb]
 \includegraphics[height=40mm]{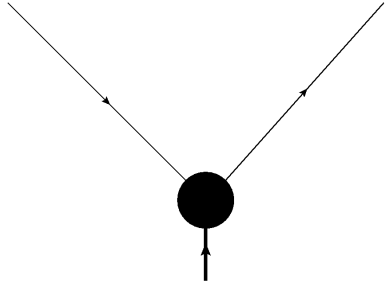}
 \caption{Tree  contribution at LO  to the pion QPDA.}
  \label{pda0}
\end{figure}

\begin{figure}[!htb]
 \includegraphics[height=50mm]{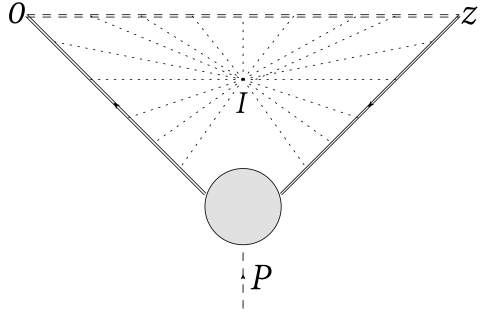}
 \caption{Star contribution at LO to the pion QPDA.}
  \label{pcac1}
\end{figure}

\subsection{QPDA  and PDA at LO}

\subsubsection{Zero mode contribution}

The first contribution in (\ref{X7X})  can be readily evaluated in singular gauge. It is solely due to the zero
modes.
We note that for $k_4=0$ a pair of poles satisfying $k_\perp^2+x^2P_z^2=\frac 14 E_\pi^2$
pinch the real $k_4$-integration line.  To address the pinch, we rotate to Minkowski space $k_4\rightarrow ik_4$,
shift  $k_4\rightarrow k_4+(x-\frac 12)P_z$ and carry the $k_z$-integration to have

\be
\label{PHI02}
\tilde{\phi}^0_{\pi}(x)\approx\lim_{P_z\rightarrow \infty}
\frac {-4iN_c}{f^2_\pi}\int \frac{dk_4d^2k_\perp}{(2\pi)^4}
\big(M(y_1)M(y_2)\big)^{\frac 12}\bigg(\frac{\bar xM(y_1)+xM(y_2)}{y_1^2y_2^2}\bigg)
\ee
with

\be
\label{13X}
&&y^2_1=-k_4(k_4+2xP_z)+k_\perp^2-i\epsilon\nonumber\\
&&y^2_2=-k_4(k_4-2\bar x P_z)+k_\perp^2-i\epsilon
\ee
The integrand in (\ref{PHI02}) involves massless poles and also square-root branch points through
the running mass (see below). 
The  $k_4$-integration  can be  carried by contour integration.
The poles are located at

\be
\label{15X}
&&k_{4\pm}=-xP_z\pm\sqrt{x^2P_z^2+k_\perp^2-i\epsilon}\nonumber\\
&&\bar k_{4\pm}=+\bar xP_z\pm\sqrt{\bar x^2P_z^2+k_\perp^2-i\epsilon}
\ee
The pair $k_{4-}, \bar k_{4+}$ moves to infinity at large momentum $P_z\rightarrow \infty$
and will be ignored (their contribution is exponentially small),  while the pair

\be
\label{16X}
&&k_{4+}\approx \frac{k_\perp^2-i\epsilon}{2xP_z}\nonumber\\
&&\bar k_{4-}\approx  \frac{k_\perp^2-i\epsilon}{-2\bar xP_z}
\ee
approaches the real-axis, on opposite sides for $x\bar x\geq 0$ and the same sides for $x\bar x<0$. In the absence of the cuts,
the QPDA has support only for $x\bar x\geq 0$ after pole closing. To proceed further, we need to address  the cuts.

\subsubsection{Unmodified effective mass at large $P_z$}

In singular gauge, the running mass $M(y_{1,2})$ at LO in (\ref{4X1})  is given in terms of modified Bessel functions $I,K$
(\ref{4X11}). When expressed in integral form, $I,K$ exhibit  $({y^2_{1,2}})^{\frac 12}$ branch points.  Note  that the branch points
are very explicit in regular gauge with (\ref{4X11X})

 \be
 M(y_{1,2})\approx e^{-2\rho(y^2_{1,2})^{\frac 12}}\nonumber
 \ee
We choose the branch-cut for the square-root function $({(a-k_4)(b+k_4)})^{\frac 12}$ to be along the negative imaginary axis such that at large $k_4>0$ the value of the square-root equals $-i$. The contour deformation of the $k_4$-integral into the upper half-plane, guarantees the positivity of the real part 
of  $y_{1,2}$ and thus the decay of $M(y_{1,2})$ asymptotically.  
For $x\bar x>0$ the contribution from the poles is purely real, while for $x\bar x <0$ their
contribution is complex. To ensure $x\leftrightarrow \bar x$ symmetry of the PDA after contour integration, the branch cuts have to be
arranged  symmetrically  for the x- and $\bar{\rm x}$-contributions in (\ref{PHI02}). With this in mind,
the result for the pion distribution amplitude (PDA)  at LO is ($k_\perp\geq M(0)$)


\be
\label{PHIZERO}
{\phi}^0_{\pi}(x)\approx\frac{2N_cM^2(0)}{f_\pi^2}\int\frac{d^2k_\perp}{(2\pi)^3}\frac 1{k_\perp ^2}
\bigg(\theta(x\bar x)\bigg(\bar xM_\perp^{\frac 12}+xM_{\underline\perp}^{\frac 12}\bigg)
+\theta(-\bar x)\,\bar x\,\bigg(M^{\frac 12}_{\perp }-\overline{M}^{\frac 32}_{\underline\perp }\bigg)
+\theta(-x)\,x\,\bigg(M^{\frac 12}_{\underline\perp }-\overline{M}^{\frac 32}_{\perp }\bigg)\bigg)
\ee
with

\begin{eqnarray}
\label{MASSPERP}
M_\perp=\frac{M(k_\perp/\sqrt{x})}{M(0)}\,\qquad\,  M_{\underline \perp}=\frac{M(k_\perp/\sqrt{\bar x})}{M(0)}\,\qquad\,
\overline{M}_\perp=\frac{M(-ik_\perp/\sqrt{x})}{M(0)}\,\qquad\,  \overline{M}_{\underline \perp}=\frac{M(-ik_\perp/\sqrt{\bar x})}{M(0)}
\end{eqnarray}
followed by the replacement $M(0)\rightarrow M(k)$ in the pion decay constant (\ref{FPI2}), to 
guarantee the normalization (\ref{X8}).
In the physical region $x\bar{x}>0$, (\ref{PHIZERO}) can be evaluated in closed form  since the integrand is a total derivative,

\be
\label{PHIZERO2}
{\phi}^0_{\pi}(x)\approx  \frac{2N_cM^2(0)}{(2\pi)^2 f_\pi^2}\left(\bar{x}F\left( \frac{\rho M(0)}{2\sqrt{x}}\right)+x F\left( \frac{\rho M(0)}{2\sqrt{\bar{x}}}\right)\right)
\rightarrow \frac {1}{{\rm ln}\bigg(\frac C{\rho M(0)}\bigg)}\,\left(\bar{x}F\left( \frac{\rho M(0)}{2\sqrt{x}}\right)+x F\left( \frac{\rho M(0)}{2\sqrt{\bar{x}}}\right)\right)
\ee
The right-most relation follows from the leading logarithm approximation for the pion decay constant, with $\rho M(0)\sim \alpha$ and

\be
\label{FZIK}
F(z)\equiv I_1(z)K_1(z)-I_0(z)K_0(z)
\ee
The infrared sensitivity of the PDA follows from the enforcement of the power counting as we noted earlier.
It matches the infrared sensitivity of the squared pion decay constant as given in (\ref{FPI2}),  and cancels
in the ratio after regulation $k_\perp\geq M(0)$ as we indicated earlier. To logarithmic accuracy, the PDA  simplifies to

\be
\phi_\pi^0(x)\rightarrow \theta (x\bar x)
\ee
with support only in the physical range and unit normalization.
In this deep infrared regime, the pion is
composed democratically of  partonic quarks in the range $0\leq x\leq 1$ including the end points.

For finite size instantons, the form factors $M_{\perp,\underline {\perp}}$ cause the PDA to
vanish at the end points $x=0,1$ as initially noted in~\cite{LAW1}, but otherwise develops spurious contributions in the non-physical region $x\bar x<0$
with real and imaginary parts. We recall that in the physical region with
 $x\bar x\geq 0$, the running mass involves a real combination of the modified Bessel functions $I,K$  as in (\ref{4X11}), and
a complex combination of the cylindrical Bessel functions $J,N$  for $x\bar x<0$ in the unphysical regions where momentum 
is conserved ($P_z=xP_z+\bar xP_z$) but energy is not ($|P_z|\neq |xP_z| +|\bar xP_z|$).

Current lattice simulations of the quasi-parton distributions~\cite{LATTICE} exhibit finite contributions outside the physically allowed x-support.
However,  they are vanishingly  small at large momentum $P_z$.  These spurious contributions relate to the transversality of the pion distribution
in the QCD instanton vacuum.
They do not arise in the $1/N_c$ analysis in two-dimensions~\cite{QCD2}. 
We now show how to remove them  approximately, without
affecting the power counting in $\alpha$ at LO, and therefore gauge and chiral symmetry.

\subsubsection{Modified effective mass at large $P_z$}

At large $P_z$, an approximative  way to eliminate the spurious contributions without 
affecting the power counting in $\alpha$, is through the substitution $M(y)\rightarrow M(k_\perp)$, which removes explicit $k_4$ dependence at the integrand level.
It  is cut-free and  restricts  the final $k_\perp$-integration to the expected physical range $M(0)\leq k_\perp \leq 1/\rho$.  Unfortunately, 
this substitution fails at the end-points  $x, \bar x=0$. To see this, we recall that for fixed $k_\perp$, the contribution to the QPDA follows 
from each of the two poles in  (\ref{15X}-\ref{16X}) with at large $P_z$

\be
\label{PAIR}
\bigg(y_1^2=0:\,\,\,  y_2^2=2k_4P_z\approx   \frac{k_\perp^2}x\bigg)\qquad{\rm and}\qquad 
\bigg(y_2^2=0:\,\,\,  y_1^2=-2k_4P_z\approx  \frac{k_\perp^2}{\bar x}\bigg)
\ee
When a quark (antiquark) goes on mass shell the anti-quark (quark) virtuality becomes parametrically large at the end points $x, \bar x=0$.
Say  $x,\bar x\approx \alpha^2\ll 1$ at the end-points, then 
the $k_\perp$-integration range at each of the pole is  vanishingly small with $M(0)\leq k_\perp\approx \alpha/\rho\approx M(0)$, 
causing the PDA to vanish. In contrast, when $x, \bar x\approx \alpha^0$ away from the end-points, the $k_\perp$-integration range is large with
$M(0)\leq k_\perp \leq 1/\rho$ in line with the leading  logarithmic approximation and power counting in (\ref{PHIZERO2}).

A simple modification  of the  induced effective quark mass (\ref{MASSZ}) at 
large $P_z$, that enforces these observations without upsetting the power counting  in $\alpha$,  that is commensurate with
(\ref{PAIR})  with manifest $x\leftrightarrow \bar x$ symmetry and free of  spurious contributions, is

\be
\label{SUBX}
M(y)\rightarrow M\bigg(\frac{k_\perp}{\lambda_\pi\sqrt{|x\bar x|}}\bigg)
=M(0)\bigg(\left|z\left(I_0K_0-I_1K_1\right)^\prime\right|^2\bigg)_{z=\frac{\rho k_\perp}{2\lambda_\pi\sqrt{|x\bar x|}}}
\equiv M(0)\bigg(\left|z F^\prime(z)\right|^2\bigg)_{z=\frac{\rho k_\perp}{2\lambda_\pi\sqrt{|x\bar x|}}}
\ee
where $\lambda_\pi\approx \alpha^0$ is a parameter of order 1, which is fixed by normalizing the PDA.
We note that the PDA is normalized in power counting at LO for the unmodified effective mass.
With this in mind, the closed-form PDA at LO following from the large $P_z$ limit is ($k_\perp\geq M(0)$)

\be
\label{PHIZERO2X}
\phi^0_{\pi}(x)\rightarrow \frac{2N_c}{f_\pi^2}\int\frac{d^2k_\perp}{(2\pi)^3}\frac {\theta( x \bar x)}{k_\perp ^2}
{M^2(k_\perp/\lambda_\pi\sqrt{x\bar x})}\rightarrow \frac{\theta(x\bar x)}{{\rm ln}\bigg(\frac C{\rho M(0)}\bigg)}
\int_{\rho M(0)/2\lambda\sqrt{x\bar x }}^\infty dz\,z^3F^{\prime\,4}(z)
\ee
 (\ref{PHIZERO2X})  is similar to (\ref{PHIZERO2}), but with  no spurious contributions!
For $x,\bar x\approx \alpha^2$, the effective quark mass in (\ref{PHIZERO2X}) is probed at virtualities larger than $1/\rho$,  which is 
still justified by noting  the agreement of the effective quark  mass with the lattice data at  large momenta  in Fig.~2.

\subsubsection{Non-zero mode contribution}

The non-zero mode contributions in (\ref{X7X}) do not vanish at finite $P_z$, but are in general
small due to the fact that at short distances $G_I\approx S_0$ or $\delta G_I\approx 0$ (UV limit),
a standard approximation  in the random instanton model. For an estimate of their contribution beyond,
we may use the Born approximation (\ref{BORN}) in (\ref{X7X}). A close inspection shows that
the ensuing color-spin traces are short of the binary pole structure $1/(y_1^2y_2^2)$ which is required
for: 1/ a finite contribution as $P_z\rightarrow\infty$; 2/ a finite contribution for $x\bar x\geq 0$.
In this approximation, the non-zero modes do not contribute to the PDA as $P_z\rightarrow\infty$.

A more explicit evaluation of the non-zero modes in (\ref{X7X}) follows from the observation that 
after analytical continuation the external quark lines are put on mass-shell.
The ensuing contribution to (\ref{X7X}) can be worked out in closed form.  Using the modified cutoff, 
and the definitions of the mass-shell conditions in Appendix E, a lengthy calculation gives

\be
\label{ONSHELL}
\phi_\pi^{\slashed{0}}(x)\approx \lim_{P_z\rightarrow \infty}\frac{\alpha}{f_\pi^2}\frac{(\sqrt{2}\pi \rho)^2}{\sigma_{00}P_z}\int\frac{d^2k_\perp}{(2\pi)^3}{M(k_\perp/\lambda_\pi\sqrt{x\bar x})}
\,{\rm Tr}\bigg({\sigma}^z\bigg(\bigg(\overline{\mathbb F}(P, k_2)-\overline{\mathbb F}(P, k_1)\bigg)-\bigg(\mathbb F(P, k_2)-\mathbb F(P, k_1)\bigg)\bigg)\bigg)
\ee
Here $k_{1,2}=k\pm P/2$ with $k_{1,2}^2\approx 0$ in the large $P_z$ limit and $P^2=0$ on mass shell.
The first contribution is from the instanton and the second contribution from the anti-instanton in the bracket. The form factors are

\be
\label{ONSHELL1}
&&\mathbb F(P,p)=\frac{\sigma_z\overline{p}+p\sigma_z}{2p\cdot P}\,f(\rho\sqrt{P^2})+
\bigg(\frac{\sigma_z(\overline P+\overline p)+(P+p)\sigma_z}{(P+p)^2}-\frac{\sigma_z\overline p+p\sigma_z}{2p\cdot P}\bigg)\,
f(\rho\sqrt{(p+P)^2})\nonumber\\
&&\overline{\mathbb F}(P,p)=\frac{\sigma_z{p}+\overline{p}\sigma_z}{2p\cdot P}\,f(\rho\sqrt{P^2})+
\bigg(\frac{\sigma_z(P+p)+(\overline P+\overline p)\sigma_z}{(\overline P+\overline p)^2}-\frac{\sigma_z p+\overline p\sigma_z}{2p\cdot P}\bigg)\,
f(\rho\sqrt{(p+P)^2})
\ee
with $f(z)=zK_1(z)-1$. 
Throughout, the Weyl notation $p=p_\mu\sigma^\mu$, $\overline{p}=p_\mu \overline{\sigma}^\mu$ etc. is used
with $\sigma^\mu=(1,\vec \sigma)$ and $\overline{\sigma}^\mu=(1,-\vec\sigma)$. The finite contribution in (\ref{ONSHELL1}) when inserted in
(\ref{ONSHELL}) cancels out. The non-zero mode contribution (\ref{ONSHELL}) to the PDA vanishes at LO.

\subsubsection{Massive Pion and Kaon}

The explicit breaking of chiral symmetry by light quark masses $u,d,s$ is understood in the QCD instanton vacuum, with the masses
for the pion and kaon  in LO obeying the GOR relation~\cite{DP,IQCD,MF}. In our case, this can be explicitly checked to hold in power counting.
For a finite current mass $m$, a rerun of the arguments leading to the effective quark mass in (\ref{4X1})  yields

\be
\label{MASSM}
M(p,m)=\frac{\alpha\,|p\varphi^{\prime }(p)|^2}{\bigg(2||q\varphi^{\prime 2}||^2+\frac{m^2}{4\alpha^2}\bigg)^{\frac 12}+\frac m{2\alpha}}+m
\equiv \frac{M(p,0)}{\bigg(1+\xi^2\bigg)^{\frac 12}+\xi}+m
\ee
with the mass parameter

\be
\xi=\frac{mM(0,0)\rho^2}{8\pi^2\kappa}
\ee
For light quarks $M(0)\equiv M(0,0)\approx 386$ MeV and $\rho\approx 1/(631 {\rm MeV})$. For massive quarks we find $M(0,5)\approx 383.7 \,\textrm{MeV}\approx M(0)$ and $M(0,150)\approx 372.6 \,\textrm{MeV} \approx M(0)$
for the up-down and strange quarks respectively. The effective quark mass is almost unchanged. As a result, the pseudoscalar decay constant for the pion and
kaon are about the same at LO for massive quarks. 

For the massive case, the integral equation (\ref{16}) for the pseudoscalar meson vertex holds  with the substitution

\be
\sigma_{00}\rightarrow {\bigg(2||q\varphi^{\prime 2}||^2+\frac{m^2}{4\alpha^2}\bigg)^{\frac 12}+\frac m{2\alpha}}\approx 
\sqrt{2}||q\varphi^{\prime 2}||+\frac m{2\alpha}
\ee
As a result the mass-shell vertex (\ref{15X0}) at LO changes to

\bea
\label{O5X}
\bigg(iO_5(P,k)\bigg)_{P^2\approx -m_P^2}\approx\frac{\sqrt{N_c}}{f_P} \sqrt{M(k)}\left(\frac{i\gamma_5}{P^2+m_P^2}\right)\sqrt{M(k-P)}
\eea
with  $f_P\approx f_\pi$ and $m_P^2\approx i2m\left<\psi^\dagger \psi\right>$ as expected.
 Note that in (\ref{O5X}) both $f_P$ and $M(k)$ are found to be {\it unaffected} by the current mass $m$
 at the meson pole at LO. The latter only shifts the meson  pole in agreement with the GOR relation.

The  ensuing PDA for massive pseudoscalars simplifies to LO

\be
\label{PHIZEROMASS}
\phi^0_{P}(x)\approx&&\frac{2N_cM^2(0)}{f_P^2}\int\frac{d^2k_\perp}{(2\pi)^3}\frac 1{k_\perp ^2-x\bar x m_P^2}\nonumber\\
&&\times\bigg(\theta(x\bar x)\bigg(\bar xM_\perp+xM_{\underline\perp}\bigg)
+\theta(-\bar x)\,\bar x\,\bigg(M_{\perp }-M_{\underline\perp }\bigg)
+\theta(-x)\,x\,\bigg(M_{\underline\perp }-M_{\perp }\bigg)\bigg)\nonumber\\
\ee
with the same cutoff $k_\perp\geq M(0)$ for light quarks $u,d,s$.
For comparison, the result for the modified effective quark mass (\ref{SUBX}) is

\be
\label{PHIZERO3X}
\phi^0_{P}(x)\rightarrow\frac{2N_c}{f_P^2}\int\frac{d^2k_\perp}{(2\pi)^3}\frac {\theta( x \bar x)}{k_\perp ^2-x\bar x m_P^2}
{M^2(k_\perp/\lambda_P\sqrt{x\bar x})}
\ee

In Fig.~\ref{fig_pda} we show the  pion PDA  (\ref{PHIZERO3X}) at LO 
for varying $\rho$ but fixed $M(0)=386$ MeV (solid curves) in comparison to the  asymptotic result of $6x\bar x$~\cite{AS}
(dashed curve). We  have set $f_\pi=93$ MeV and $m_\pi=135\,\textrm{MeV}$ and  fixed $\lambda_\pi = 3.41894$ for the overall normalization of the PDA
with the modified effective mass. (No  such a modification is needed for the unmodified effective quark mass). The result at this low renormalization scale $Q_0=1/\rho$ is remarkably close to the QCD asymptotic result of $6x\bar x$~\cite{AS}.  The single $q\bar q$-component  of the pion wavefunction is well described 
in the random instanton vacuum (RIV) in the planar approximation at LO. Since the constituent mass 
$M(0)\approx M(0,5)\approx M(0, 150)$ is almost unchanged for $u,d,s$, the kaon PDA   is almost undistinguishable from the pion PDA at LO.

Our result for the pion PDA at LO is similar  to the one obtained originally  in~\cite{LAW1} using time-like arguments
with  a modified dipole effective quark mass with very different analytical properties. 
It is overall analogous to the one derived from modified holographic models~\cite{BROD}.
As  $\rho\rightarrow 0$, and the cutoff is removed, the pion PDA 
asymptotes the middle-solid-red curve in Fig.~\ref{fig_pda} which is close to the normalized step function $\theta (x\bar x)$. The same result was  noted for chiral quark models with point interactions~\cite{BASIS,BRO}, and some bound-state resummations~\cite{M3}. 

In Fig.~\ref{fig_pda2} we compare our result for the pion PDA  shown in red-solid line (RIV)  to the recently generated pion PDA 
blue-wide-band, using lattice simulations using the large momentum effective theory (LaMET)~\cite{JI}. The QCD asymptotic result 
black-dashed curve is again shown for comparison.

\begin{figure}[!htb]
\centering
 \includegraphics[height=6cm]{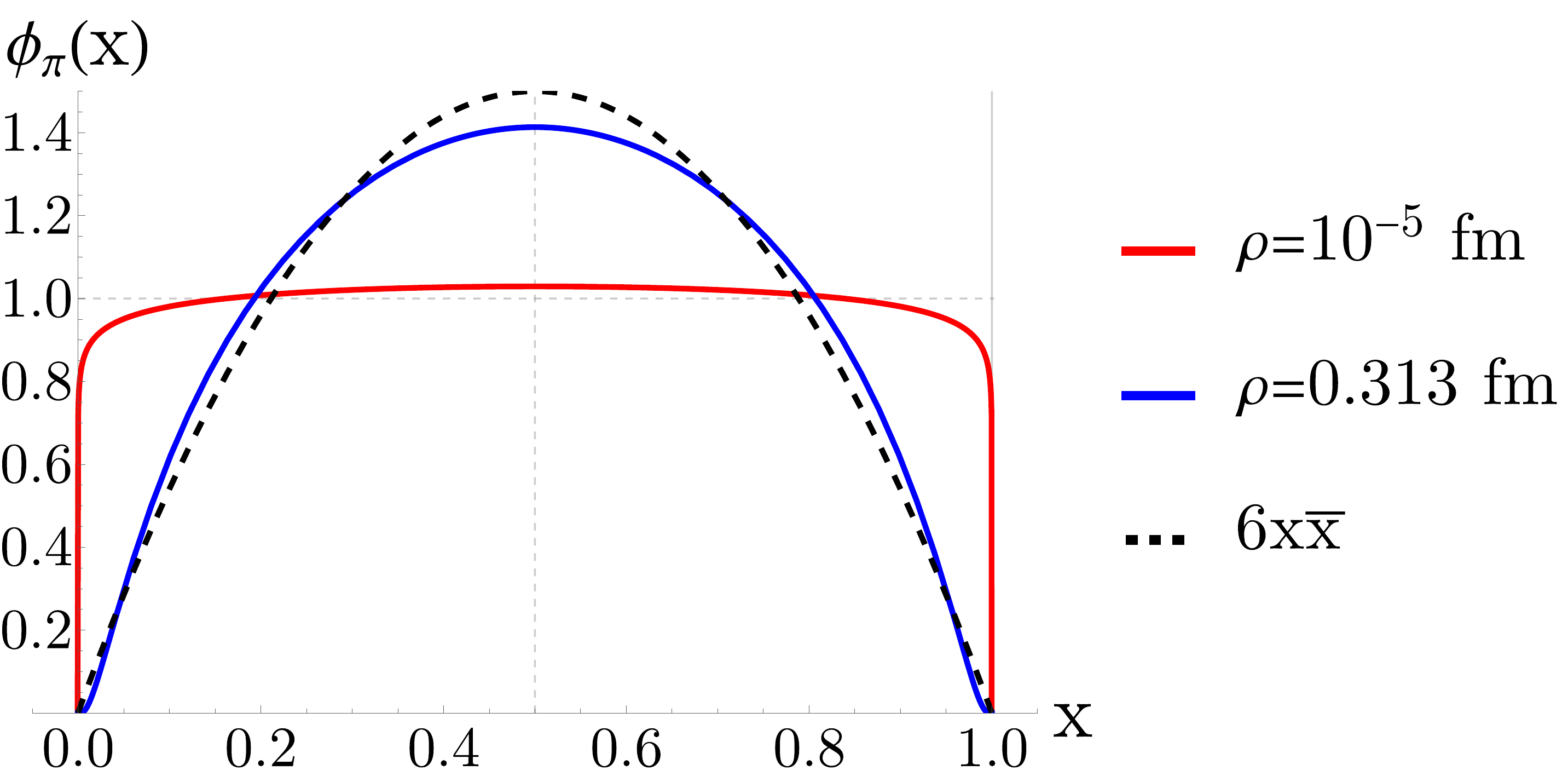}
 \caption{Pion PDA (\ref{PHIZERO3X}) for varying instanton size $\rho$ but fixed $M(0)=386$ MeV (solid curves) in comparison to the  asymptotic result of $6x\bar x$~\cite{AS}
 (dashed curve).}
  \label{fig_pda}
\end{figure}

\begin{figure}[!htb]
\centering
 \includegraphics[height=6cm]{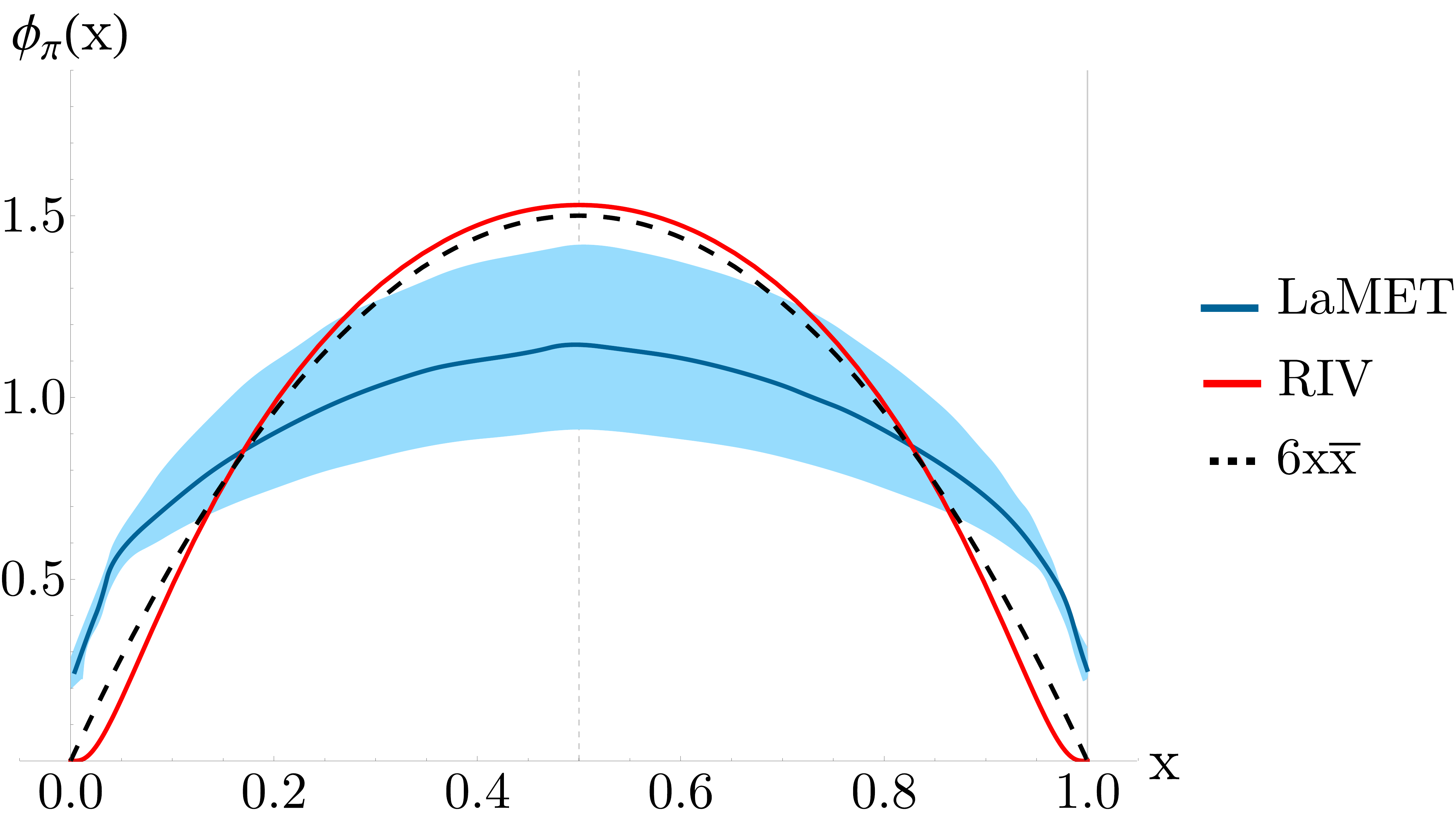}
 \caption{Pion PDA  from the random instanton model (RIV, solid-red-curve) (\ref{PHIZERO3X}), 
 asymptotic QCD (dashed-black-curve)~\cite{AS} in comparison to the  lattice simulations LaMET (blue-wide-band)~\cite{JI}}.
  \label{fig_pda2}
\end{figure}

\subsection{QCD evolution of pion PDA}

The pion PDA (\ref{PHIZERO3X}) is defined at a low renormalization scale set by the instanton size $Q_0=1/\rho=631$ GeV.  Assuming factorization, 
its form at higher  renormalization scales follows from a QCD kernel evolution equation (ERBL).  Its  closed form solution in the form of Gegenbauer polynomials was given in~\cite{AS}.  More specifically, using (\ref{PHIZERO2X}) as an initial condition, the ERBL evolved  pion PDA is~\cite{AS}

\begin{figure}[!htb]
\centering
 \includegraphics[height=6cm]{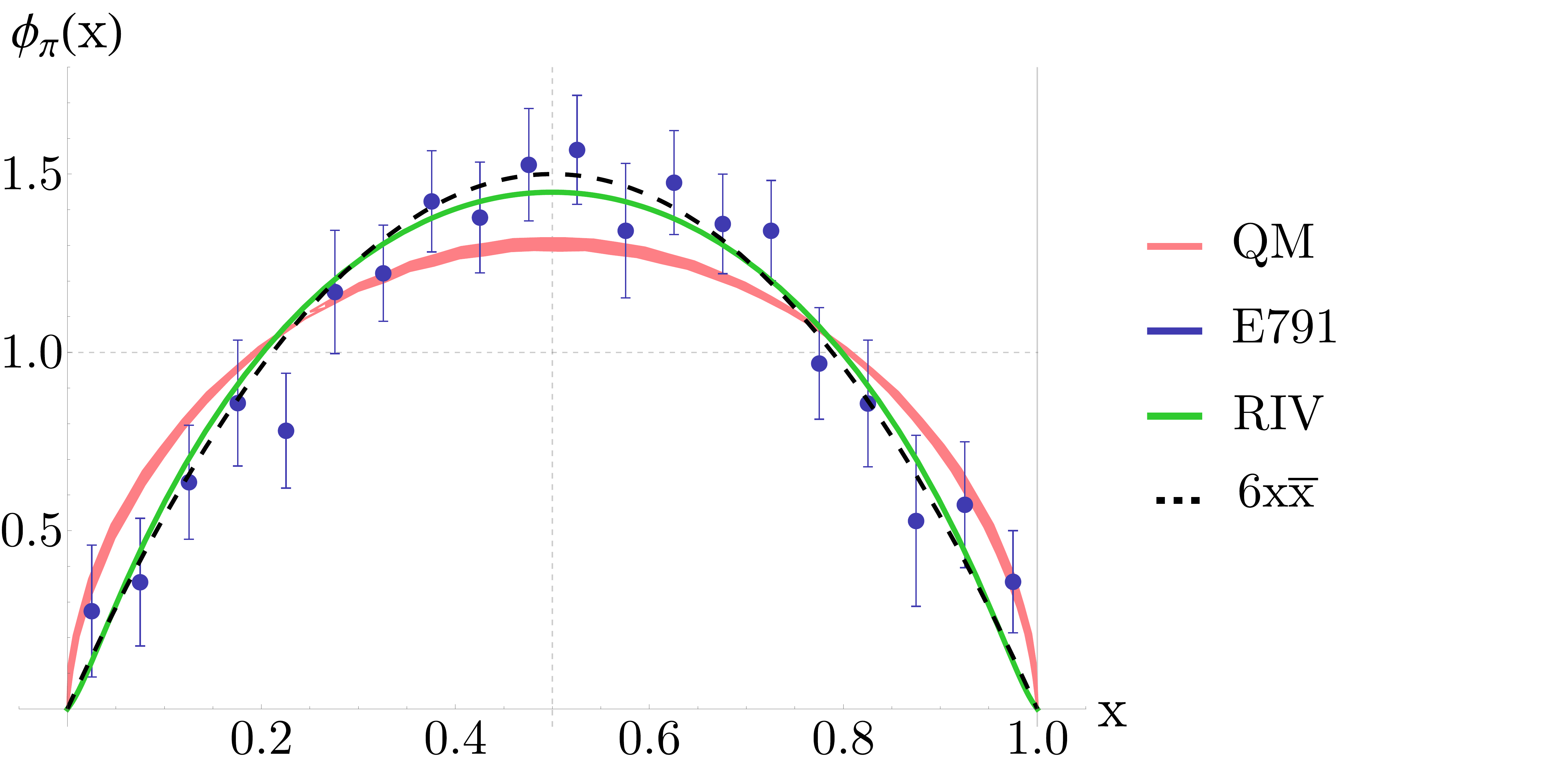}
 \caption{Pion PDA  from the random instanton model evolved to $Q=2\,\textrm{GeV}$ (RIV, solid-red-curve) (\ref{PHIZERO2X}), the quark model QM@2GeV (QM, solid-pink-band)~\cite{BRO}, 
 asymptotic QCD (dashed-black-curve)~\cite{AS} in comparison to the E791 dijet data~\cite{E791}}.
  \label{fig_pda1}
\end{figure}

\be
\phi_\pi(x,Q)=6x\bar x\sum_{n-even}a_n(Q_0)\bigg(\frac{\alpha_s(Q)}{\alpha_s(Q_0)}\bigg)^{\gamma_n/\beta_0}\,C_n^{\frac 32}(x-\bar x)
\label{GE1}
\ee
with the initial coefficients

\be
\label{GE2}
a_n(Q_0)=\frac{2(2n+3)}{3(n+1)(n+2)}\int_0^1dy\,C_n^{\frac 32}(y-\bar y)\,\phi^0_\pi(y)
\ee
Here $\alpha_s(Q)=4\pi/\beta_0{\rm ln}(Q^2/\Lambda^2)$ is the one-loop running QCD coupling with $\beta_0=\frac {11}3 N_c-\frac 23 N_f$
and $\Lambda=250$ MeV ($\overline{\rm MS}$-scheme). The  $\gamma_n$ are pertinent  anomalous dimensions

\be
\label{ANOMALOUSDIM}
\gamma_n=C_F\bigg(1+4\sum_{k=2}^{n+1}\frac 1k-\frac 2{(n+1)(n+2)}\bigg)
\ee
with the Casimir $C_F=(N_c^2-1)/2N_c$.
Since $\gamma_0=1$ and $\gamma_n>0$, it follows that (\ref{GE1}) asymptotes $6x\bar x$ with $a_0(Q_0)=1$ for $Q\rightarrow\infty$ as 
illustrated in Fig.~\ref{fig_pda}.

In Fig.~\ref{fig_pda1} we show the ERBL evolved pion PDA  (\ref{PHIZERO2X}) at $Q=2$ GeV as a green-solid curve (RIV), which is in good agreement with
the empirical pion PDA  blue-data points extracted from dijet data by the E791 collaboration~\cite{E791} at the same scale. For comparison, we also show the chiral quark model evolved PDA to $Q=2$ GeV as a solid-pink-band (QM)~\cite{BRO} and the asymptotic QCD result~\cite{AS}. Again, since $M(0)$ does not change
much for massive pions and kaons, the ERBL evolved kaon PDA is undistinguishable from its evolved pion counterpart at LO in the present analysis.

\section{Pion quasi-parton distribution function}

In this section we show how to re-sum the planar contributions to the three-point functions in general.
We then apply the results to the derivation of the pion quasi-parton distribution function to LO.
Since this distribution obeys charge and momentum sum rules, the enforcement of the gauge and
chiral symmetry through the Ward identity  is  needed.

\subsection{Three-point function}

The quasi-parton distributions involve 3-point functions with one of the source point-split. In the planar approximation, their
construction follows a similar reasoning as the one developed earlier.  For that, consider the general
3-point function

\be
\label{39}
\left<O_1 O_3 O_2\right>
\ee
where the $O^\prime$s are resummed and colorless local or quasi-local fermionic bilinears defined as

\be
\label{40}
O_{ab}={\rm Tr}_C\left( S_{\gamma \beta}O_{\beta \alpha}S_{\alpha \delta} T_{a\gamma;\delta b} \right)
\ee
and are spin-flavor valued in general.
In the planar approximation, the leading contributions to (\ref{39}) are

\be
\label{41}
\left<O_1 O_3 O_2\right> =
{\rm Tr}_C(O_3S O_2 SO_1 S)+\frac{N}{2N_cV }\int_{I+\bar I} dz_I
{\rm Tr}_C\left(O_3S (-\Sigma_I)S O_2 S(-\Sigma_I) S O_1 S(-\Sigma_I)S\right)
\ee
The first contribution sums up all planar diagrams with no common instanton to the three quark lines
as illustrated in Fig.~\ref{pcac4}.
The second contribution corresponds to the planar contributions with one instanton shared by the three
quark lines.  Planarity implies that only {one} instanton is commonly shared by the three quark lines  as shown in Fig.~\ref{pcac2}.
For a finite gauge link $[z_-,z_+]$ there is an additional contribution shown in Fig.~\ref{pcac5} with $I,J$ referring to a double summation over
distinct instantons (anti-instantons). It is readily seen that this contribution reduces to that shown in Fig.~\ref{pcac2} when the gauge link is 1,
so it will be ignored. The direct and cross contributions follow from pertinent re-routing of the momenta.
The extension of these observations to  the $n$-point functions is now straightforward.



\begin{figure}[!htb]
 \includegraphics[height=50mm]{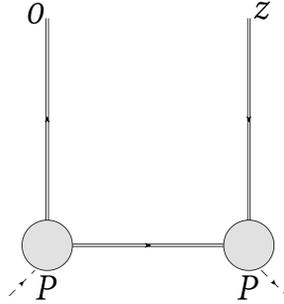}
 \caption{Tree contribution at LO  to the  pion QPDF}
  \label{pcac4}
\end{figure}

\begin{figure}[!htb]
 \includegraphics[height=50mm]{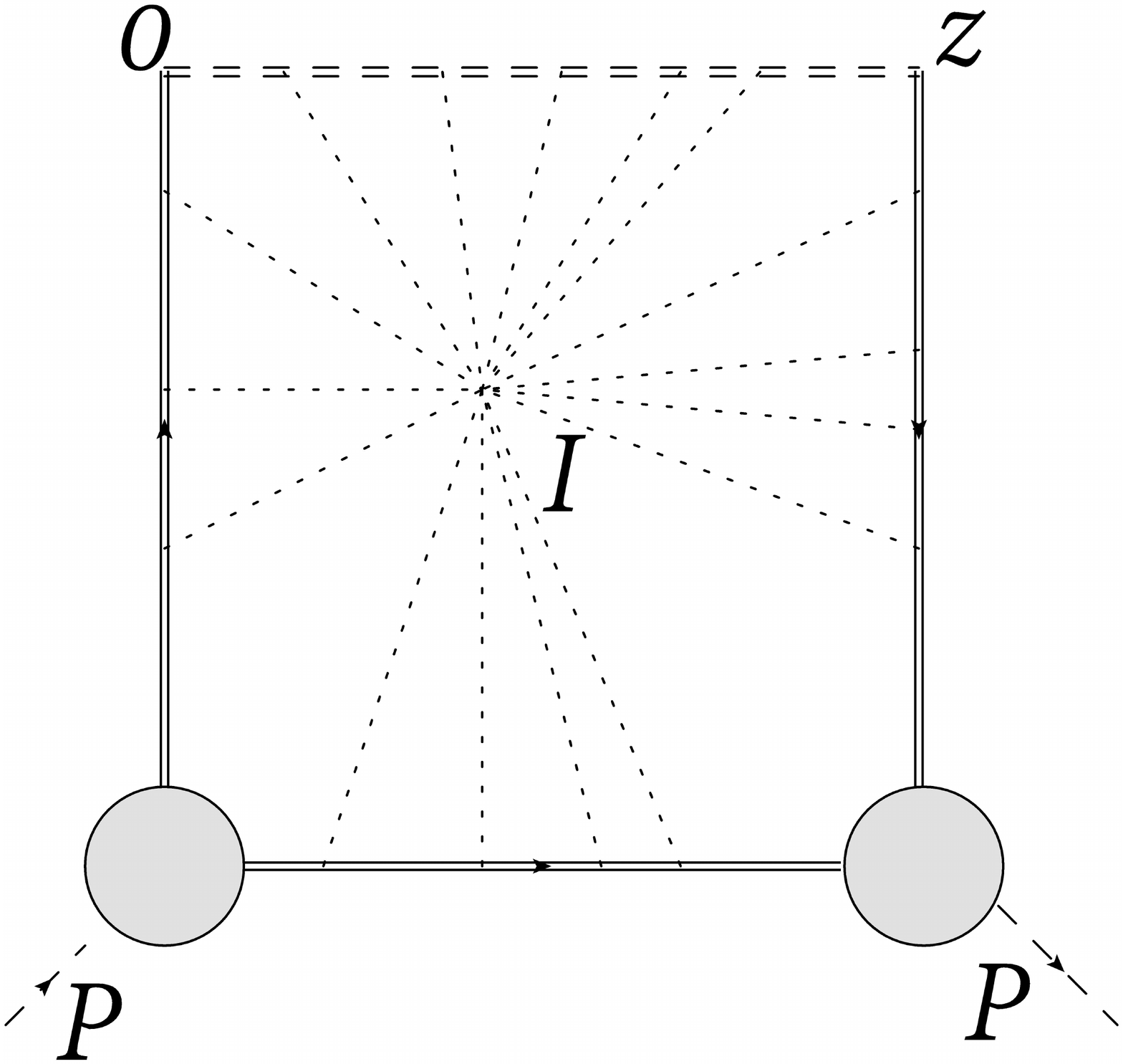}
 \caption{Star contribution at LO to the pion QPDF}
  \label{pcac2}
\end{figure}

\begin{figure}[!htb]
 \includegraphics[height=50mm]{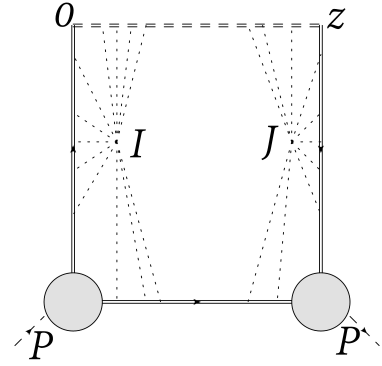}
 \caption{Split  contribution at LO to the pion QPDF}
  \label{pcac5}
\end{figure}


\subsection{Pion QPDF   and PDF at LO}

The pion quasi-distribution function (QPDF) can also be extracted from the equal-time correlator
following (\ref{1}) as suggested in~\cite{JI1}. In our case it follows by reduction using the pseudoscalar
source. Specifically,  in the chiral limit we have

\be
\label{Z1X}
&&\tilde  \psi_\pi(x, P_z)=\int \frac{dz}{2\pi}\,e^{-\frac i2 (x-\bar x)zP_z}\, \left<\pi(P)\right|{\psi}^\dagger(z_-)\gamma^z\,[z_-, z_+]\psi(z_+)\left|\pi(P)\right>\nonumber\\
&&\approx \lim_{P^2\rightarrow 0}\frac{P^4}{g_\pi^2}\int \frac{dz}{2\pi}\,e^{-\frac i2 (x-\bar x)zP_z}\,
\left<O_5(-P){\psi}^\dagger(z_-)\gamma^z\,[z_-,z_+]\,\psi(z_+)O_5(P)\right>\nonumber\\
\ee
Following the previous reasoning we may approximate the gauge link to 1 in the large $P_z$ limit.
Using the density expansion for the pseudoscalar vertex (\ref{PPNLO}) and the effective mass (\ref{NLO3})
at NLO,  we can unwind (\ref{Z1X}). The  result at LO

\be
\label{Z3X}
&&\tilde\psi_\pi(x,P_z)\approx \nonumber\\
&&
- \lim_{P^2\rightarrow 0}\frac{P^4}{g_\pi^2P_z}
\int \frac{d^4k}{(2\pi)^4} \delta \left(x-\frac 12-\frac {k_z}{P_z}\right)
{\rm Tr}_C\bigg(\gamma^z\frac 1{\slashed{k}_1}\gamma^5F_5(P,k)\frac 1{\slashed {k}_2}\gamma^5
F_5(P, k)\frac 1{\slashed{k}_1}\bigg)+ {\rm cross}\nonumber\\
&&+\lim_{P^2\rightarrow 0}\frac{P^4}{g_\pi^2\sigma_{00}^2P_z}
\int \frac{d^4k}{(2\pi)^4}\frac{d^4q}{(2\pi)^4}\frac{d^4p}{(2\pi)^4} \delta \left(x-\frac 12-\frac {k_z}{P_z}\right)
{\rm Tr}_C\bigg(\gamma^z\psi_{0I}(k_1)\psi_{0I}^\dagger(q_1)\gamma^5F_5(P, q)
\delta G_I(q_2, p_2)\gamma^5F_5(P, p)\psi_{0I}(p_1)\psi_{0I}^\dagger(k_1)\bigg)\nonumber\\
&&+\lim_{P^2\rightarrow 0}\frac{P^4}{g_\pi^2\sigma_{00}^2 P_z}
\int \frac{d^4k}{(2\pi)^4}\frac{d^4q}{(2\pi)^4}\frac{d^4p}{(2\pi)^4} \delta \left(x-\frac 12-\frac {k_z}{P_z}\right)
{\rm Tr}_C\bigg(\gamma^z\psi_{0I}(k_1)\psi_{0I}^\dagger(q_1)\gamma^5F_5(P, q)
\psi_{0I}(q_1)\psi_{0I}^\dagger(p_2)\gamma^5F_5(P, p)\delta G_I(p_1, k_1)\bigg)\nonumber\\
&&+\lim_{P^2\rightarrow 0}\frac{P^4}{g_\pi^2\sigma_{00}^2 P_z}
\int \frac{d^4k}{(2\pi)^4}\frac{d^4q}{(2\pi)^4}\frac{d^4p}{(2\pi)^4} \delta \left(x-\frac 12-\frac {k_z}{P_z}\right)
{\rm Tr}_C\bigg(\gamma^z\delta G_I(k_1, q_1)\gamma^5F_5(P, q)\psi_{0I}(q_2)\psi_{0I}^\dagger(p_2)
\gamma^5F_5(P, p)\psi_{0I}(p_1)\psi_{0I}^\dagger(k_1)\bigg)\nonumber\\
\ee
with $k_{1,2}=k\pm \frac P2$ and so on. The summation over $I, \bar I$ is subsumed.
The cross refers to the cross contributions (see below).
All contributions are of order $\alpha^0$ since $g_\pi\sim \alpha^0$ and $\sigma_{00}\sim \alpha^0$.
The first contribution involves only the zero modes. The second to fourth contributions
involve the cross contribution from the zero and non-zero modes. The latters are required
for the enforcement of the Ward identities in power counting, and
all contributions are of the same order in $\alpha$. We note that the second contribution in (\ref{Z3X})
vanishes due to a mismatch in chirality.


\subsubsection{Non-zero mode contribution}

An  explicit evaluation of the non-zero modes in (\ref{Z3X}) is involved, but follows from the observation that 
after analytical continuation the external quark lines are put on mass-shell as we noted earlier for the PDA. 
In Appendix E  the rules  for putting the instanton zero modes and non-zero mode propagator on mass shell are given.
The ensuing contribution to (\ref{Z3X}) can be worked out in closed form much like for the PDA.  Using the modified cutoff, a lengthy calculation gives

\be
\label{ONSHELL3}
\psi_\pi^{\slashed{0}}(x)\approx \lim_{P_z\rightarrow \infty}\frac{(\sqrt{2}\pi \rho)^2}{f_\pi^2\sigma_{00}^2 P_z}\int\frac{d^2k_\perp}{(2\pi)^3}{M^2(k_\perp/\lambda_\pi\sqrt{x\bar x})}
\,{\rm Tr}\bigg({\sigma}^z\bigg(2{\mathbb F}(0, k_1)+2\overline{\mathbb F}(0,k_1)\bigg)
\ee
with $k_1=k+P/2$ and $k_1^2\approx 0$ at large $P_z$. The form factors are
given in  (\ref{ONSHELL1}). They are zero for the present kinematics. The non-zero mode contribution (\ref{ONSHELL3})
vanishes. So in the large momentum limit the pion PDF at LO is dominated by the zero mode contribution which we now explicit. 

\subsubsection{Pion and Kaon PDF  at  LO and  large $P_z$}

The first contribution in (\ref{Z3X})  is  dominated by the pion pole. Inserting  (\ref{16}),
carrying the spin trace, unwinding the $k_z$-integration and analytically continuing
$k_4\rightarrow ik_4$ yield  ($k_\perp\geq M(0)$)

\be
\label{Z4X}
\tilde\psi^0_\pi(x,P_z)\approx
\frac{4iN_c}{f_\pi^2}
\int\frac{dk_4 d^2k_\perp}{(2\pi)^4}\,
\bigg(M(y_1)M(y_2)\bigg(\frac{x+\bar x}{y_1^2y_2^2}+\frac{x}{y_1^4}\bigg)+M(y_1)M(y^{\prime}_2)\bigg(-\frac{x+\bar x}{y_1^2(y^{\prime}_2)^2}+\frac{x}{y_1^4}\bigg)\bigg)
\ee
with $y_{1,2}$ given in (\ref{13X}), and $y^{\prime}_2=k+P$ in the cross contribution.  (\ref{Z4X}) can be undone by pole closing.
In the large $P_z$ limit, the cross contribution in (\ref{Z4X}) and the 
contribution $1/y_1^4$ in (\ref{Z4X}) are  subleading. Using the unmodified effective quark mass (\ref{MASSZ}), the result for the pion
 PDF at LO and large $P_z$  and in the chiral limit is 

\be
\label{PSIZERO}
\psi^0_{\pi}(x)\approx\frac{2N_cM^2(0)}{f_\pi^2}\int\frac{d^2k_\perp}{(2\pi)^3}\frac 1{k_\perp ^2}
\bigg(\theta(x\bar x)\bigg(\bar xM_\perp+xM_{\underline\perp}\bigg)
+\theta(-\bar x)\,\bar x\,\bigg(M_{\perp }-M_{\underline\perp }\bigg)
+\theta(-x)\,x\,\bigg(M_{\underline\perp }-M_{\perp }\bigg)\bigg)\nonumber\\
\ee
Note that a similar conclusion follows from  the free approximation for the non-zero modes $\delta G_I\approx 0$, or  the Born
approximation (\ref{BORN}). For comparison, the result with the modified effecttive quark mass (\ref{SUBX}) is

\be
\label{PSIZERO1}
\psi^0_{\pi}(x)\rightarrow \frac{2N_c}{f_\pi^2}\int\frac{d^2k_\perp}{(2\pi)^3}\frac {\theta( x \bar x)}{k_\perp ^2}
 {M^2(k_\perp/\lambda_\pi\sqrt{x\bar x})}\approx \phi_\pi^0(x)
\ee

Away from the chiral limit, the QPDF  involve several contributions that will be presented elsewhere. We have
checked that the PDF limit at LO simplifies. For the unmodified quark effective mass (\ref{MASSZ}) the PDF for
the f-flavor in the the P-pseudoscalar or $f/P$ is

\be
\label{PSIZEROZ}
\psi^0_{f/P}(x)\approx\frac{2N_cM^2(0)}{f_P^2}\int_{f}\frac{d^2k_\perp}{(2\pi)^3}\frac {k_\perp^2}{(k_\perp^2-x\bar x m_P^2)^2}
\bigg(\theta(x\bar x)\bigg(\bar xM_\perp+xM_{\underline\perp}\bigg)
+\theta(-\bar x)\,\bar x\,\bigg(M_{\perp }-M_{\underline\perp }\bigg)
+\theta(-x)\,x\,\bigg(M_{\underline\perp }-M_{\perp }\bigg)\bigg)\nonumber\\
\ee
while for the modified quark effective mass (\ref{SUBX}) it is

\be
\label{PSIZEROX}
\psi^0_{f/P}(x)\rightarrow\frac{2N_c}{f_P^2}\int_{f}\frac{d^2k_\perp}{(2\pi)^3}\frac {\theta( x \bar x)\,k_\perp^2}{(k_\perp ^2-x\bar x m_P^2)^2}
 {M^2(k_\perp/\lambda_P\sqrt{x\bar x})}
\ee
The $f$-integration is carried with  $k_\perp\geq M(0,m_f)$, with  $f=u,d$ for the pion and $f=u,s$
for the kaon.


\subsection{QCD evolution of pion and kaon PDF}

To compare the pion  and kaon PDF in the random instanton vacuum at the inverse instaon size scale $Q_0=1/\rho=631$ GeV, 
with the measured pion PDF at higher resolution we need to evolve the pion PDF (\ref{PSIZEROX}) 
to a higher scale using QCD evolution (DGLAP). A more appropriate evolution with a modified DGLAP kernel including small
size instanton corrections will be discussed elsewhere.  With this in mind, the one-loop DGLAP evolution of the forward (non-singlet) pseudoscalar PDF $\psi_{P}(x,t)$ is

\be
\label{DGLAP1}
\frac{d\psi_{P}(x,t)}{dt}=\frac{\alpha_s (t)}{2\pi}\int_x^1 \frac{dy}{y}P_{qq}^{(0)}\left(\frac{x}{y}\right) \psi_{P}(y,t)
\ee
with $t=\log (Q^2/\Lambda_{\textrm{QCD}}^2)$ and $P_{qq}^{(0)}(z)$ is the one-loop non-singlet splitting function

\be
\label{SPLITTING}
P_{qq}^{(0)}(z)=C_F\left[\frac{1+z^2}{(1-z)_+}+\frac{3}{2}\delta(1-z)\right]
\ee
We numerically evolve from $t$ to $t+\Delta t$ by simple forward-Euler. We sample $\psi_{P}(x,t)$ on a uniform grid in $x$, create a spline-interpolation, and evaluate the RHS of (\ref{DGLAP1}) to calculate $d\psi$. Consistency of the evolution is checked in two ways: first by verifying that the first few Mellin moments evolve according to the analytical result

\be
\label{MELLIN}
M_n(t)\equiv \int_0^1 x^n \psi_P(x,t)\,dx = M(t_0)\left(\frac{\alpha_s(t)}{\alpha_s(t_0)}\right)^{\gamma_n/\beta_0}
\ee
where $\gamma_n$ is the same as before in (\ref{ANOMALOUSDIM}), and second by we reproducing the evolution of~\cite{BRO} where the authors evolve a step-function $\psi_P(x,t_0)=\theta(x\bar x)$ from $Q_0=313\,\textrm{MeV}$ to $Q=2\,\textrm{GeV}$ with $\Lambda_{\textrm{QCD}}=226\,\textrm{MeV}$.

\begin{figure}[!htb]
\centering
 \includegraphics[height=50mm]{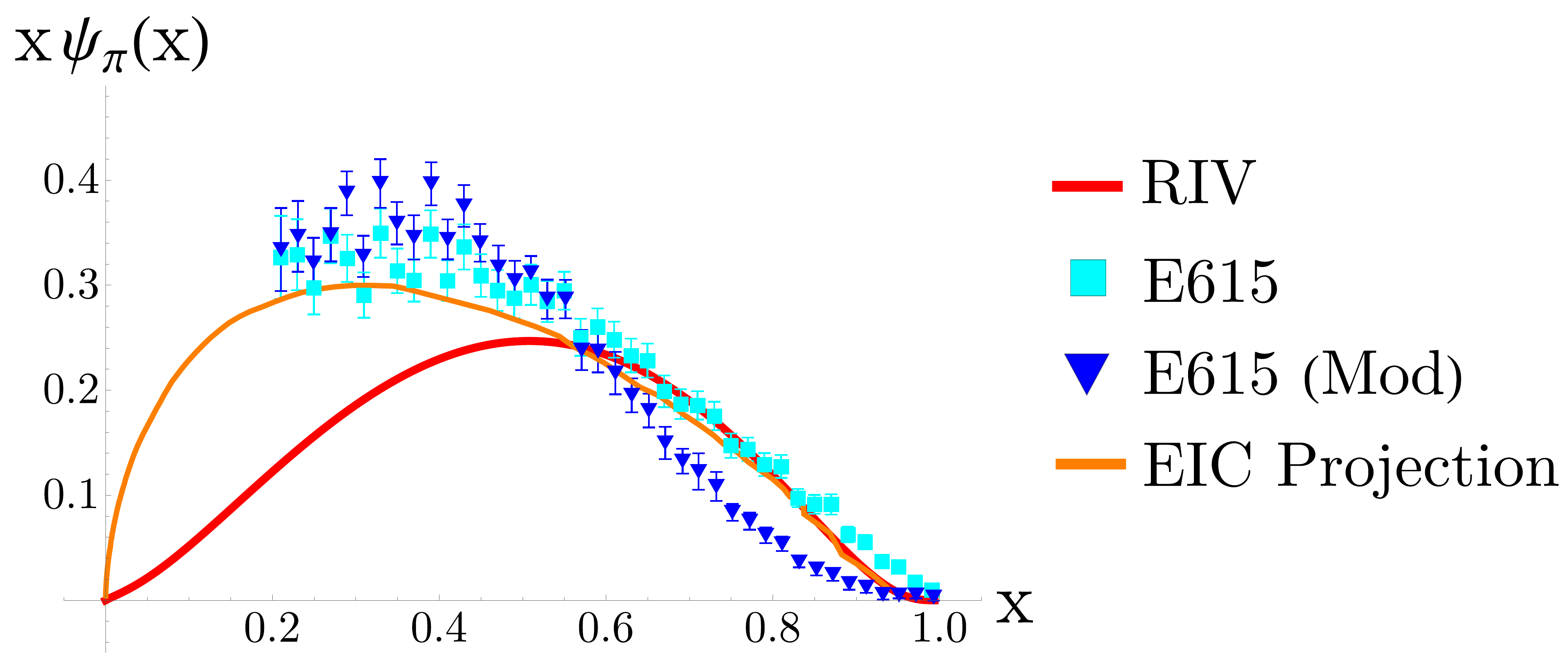}
 \caption{Pion longitudinal momentum distribution  in the QCD instanton vacuum RIV solid-red-curve (\ref{PSIZERO}), the E615 data blue-square~\cite{E615}
 and improved E615 data  inverse-blue-triangle~\cite{E615MOD}, and the EIC projection in solid-orange-curve~\cite{EICPRO}. All are evolved to $Q^2=4 \textrm{GeV}^2$.}
  \label{fig_pdf1}
\end{figure}

\begin{figure}[!htb]
\centering
 \includegraphics[height=50mm]{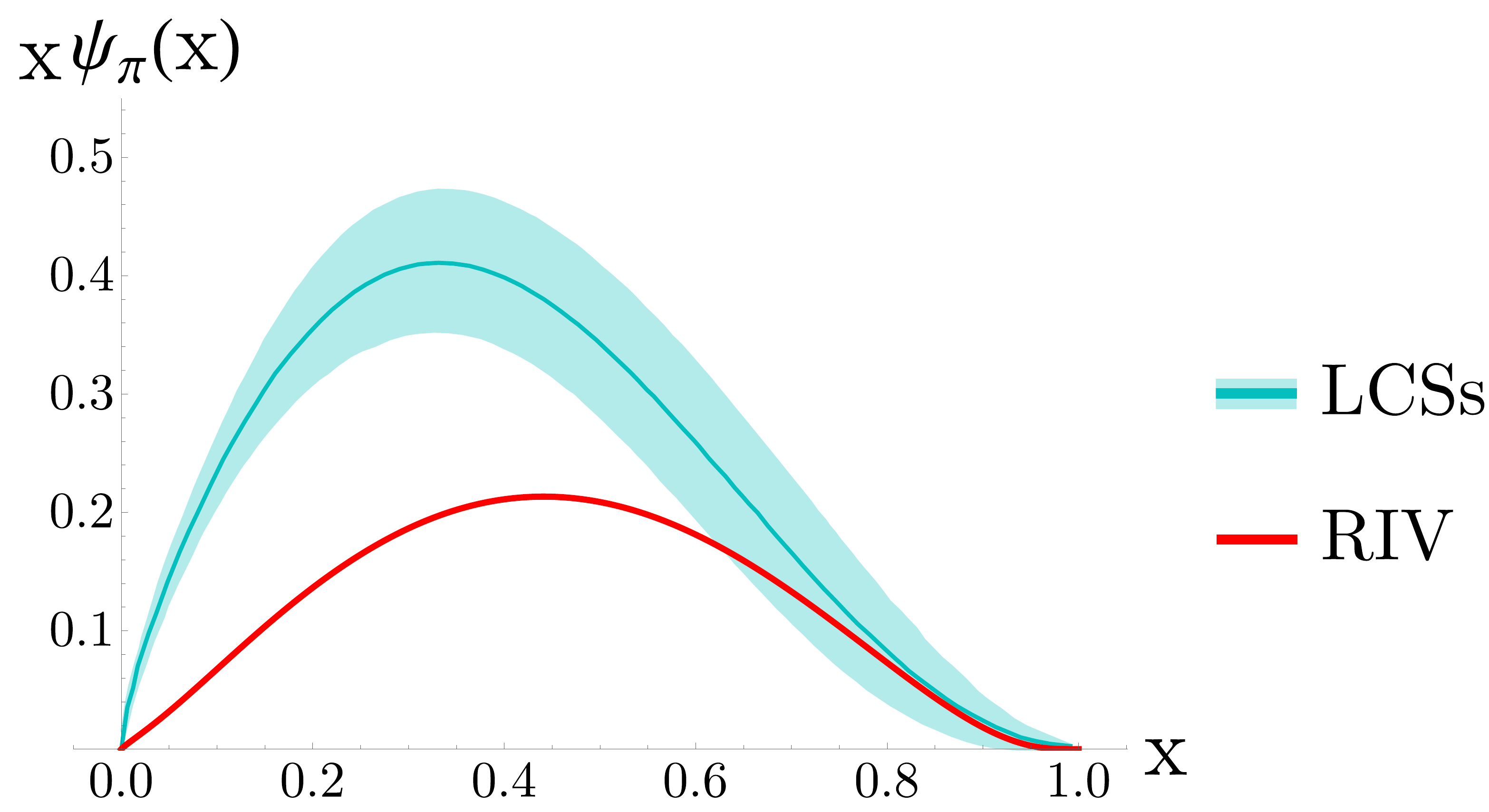}
 \caption{Pion longitudinal momentum distribution
 in the QCD instanton vacuum evolved to $Q^2=27\,\textrm{GeV}^2$ (RIV, solid-red-curve) (\ref{PSIZERO}) in comparison to the lattice results LCSs blue-wide-band at the same scale~\cite{LCS}.}
  \label{fig_pdf}
\end{figure}

In Fig.~\ref{fig_pdf1} we show the result for  the pion longitudinal  momentum distribution in our QCD random instanton vacuum (RIV) 
in solid-red-curve (\ref{PSIZERO}) evolved to $Q^2=4$ GeV$^2$. The data are from the  E615 collaboration blue-square~\cite{E615},
and the improved E615 data  inverse-blue-triangle~\cite{E615MOD}. The EIC projection is shown in solid-orange-curve~\cite{EICPRO}. 
All are evolved to the same $Q^2=4 \textrm{GeV}^2$. 
In Fig.~\ref{fig_pdf} we show the  pion longitudinal distribution in the QCD random instanton vacuum (RIV) solid-red-curve  in
comparison to recent lattice results (LCSs) as a blue-wide-band~\cite{LCS} at a higher scale $Q^2=27$ GeV$^2$.
There is good agreement at large-x, but the RIV results fall short at low-x. This maybe a shortcoming of our planar approximation 
which ignores multi-$q\overline q$ or sea contributions to the pion wavefunction at low-x. We note that for smaller size instantons,
the pion (kaon) PDF  shown in Fig.~\ref{fig_pda} flatens out. Its DGLAP evolution is more in line with the data  for all-x.
However, smaller size instantons do not support the key vacuum parameters we have established  earlier.

\subsection{Pion TMD at  LO and  large $P_z$}

Finally, we note that the integrand in (\ref{PSIZEROX}) describes the parton transverse momentum distribution (TMD) in a pseudoscalar $P$. However at this point we must recall (\ref{KPSHIFT2}) --- that our actual leading-order TMD is only obtained after shifting back $k_{\perp}^2\rightarrow k_{\perp}^2+M^2(0)$. It follows that
the TMD for the massive pion at LO is

\be
\label{GPSIZERO0}
\psi^0_{\pi}(x, k_\perp)\rightarrow \frac{2N_c}{f_\pi^2}\frac{1}{(2\pi)^3}\,
\frac {\theta( x \bar x)\,(k_\perp^2+M^2(0))}{(k_\perp ^2+M^2(0)-\bar xx m_{\pi}^2)^2}
{M^2\left(\frac{\sqrt{k_\perp ^2+M^2(0)}}{\lambda_\pi\sqrt{x\bar x}}\right)}
\ee

while the transverse spatial distribution is

\be
\label{GPSIZERO}
\psi^0_{\pi}(x, b_\perp)\rightarrow\frac{2N_c}{f_\pi^2}\int\frac{d^2k_\perp}{(2\pi)^3}\,e^{ik_\perp\cdot b_\perp}\,
\frac {\theta( x \bar x)\,k_\perp^2}{(k_\perp ^2-\bar xx m_{\pi}^2)^2}
{M^2(k_\perp/\lambda_\pi\sqrt{x\bar x})}
\ee

with $k_\perp\geq M(0)$ subsumed.  The leading logarithm contribution to the TMD  in the massless case is

\be
\label{GPSIZERO1}
\psi^0_{\pi}(x, b_\perp)\rightarrow
\frac{\theta(x\bar x)}{4{\rm ln}\bigg(\frac C{\rho M(0)}\bigg)}
\int_{\rho M(0)/2\lambda_\pi\sqrt{x\bar x }}^\infty dz\,J_0\bigg(2\sqrt{x\bar x}\frac{zb_\perp}\rho\bigg)\,z^3F^{\prime\,4}(z)
\ee
For comparison, the massless  pion TMD with the unmodified effective quark mass (\ref{MASSZ}) in the physical  region $x\bar x\geq 0$,  is

\be
\label{GPSIZERO1}
\psi^0_{\pi}(x,b_\perp)\approx \frac{N_c\rho^2M^2(0)}{2f_\pi^2}\int_{M(0)}^\infty \frac{k_\perp dk_\perp}{(2\pi)^2}\,J_0(k_\perp b_\perp)
\bigg(\frac{\overline{x}}x\,F^{\prime\,2}\bigg({z}_k=\frac{\rho k_\perp}{2\sqrt{x}}\bigg)
+\frac{{x}}{\overline x}\,F^{\prime\,2}\bigg(\overline{z}_k=\frac{\rho k_\perp}{2\sqrt{\overline x}}\bigg)\bigg)
\ee
with $F^\prime (z)$  the z-derivative of  (\ref{FZIK}).

In Fig.~\ref{fig_tmdB} we show the pion and kaon transverse spatial distributions from the QCD random 
instanton vacuum (\ref{GPSIZERO1}),  at  the low renormalization scale $Q_0=631\,\textrm{MeV}$.
The corresponding  distributions in transverse momentum space  are  also shown in Fig.~\ref{fig_tmdP} at the same  scale.

\begin{figure}[!htb]
\centering
\begin{subfigure}{0.5\textwidth}
  \centering
  \includegraphics[height=50mm]{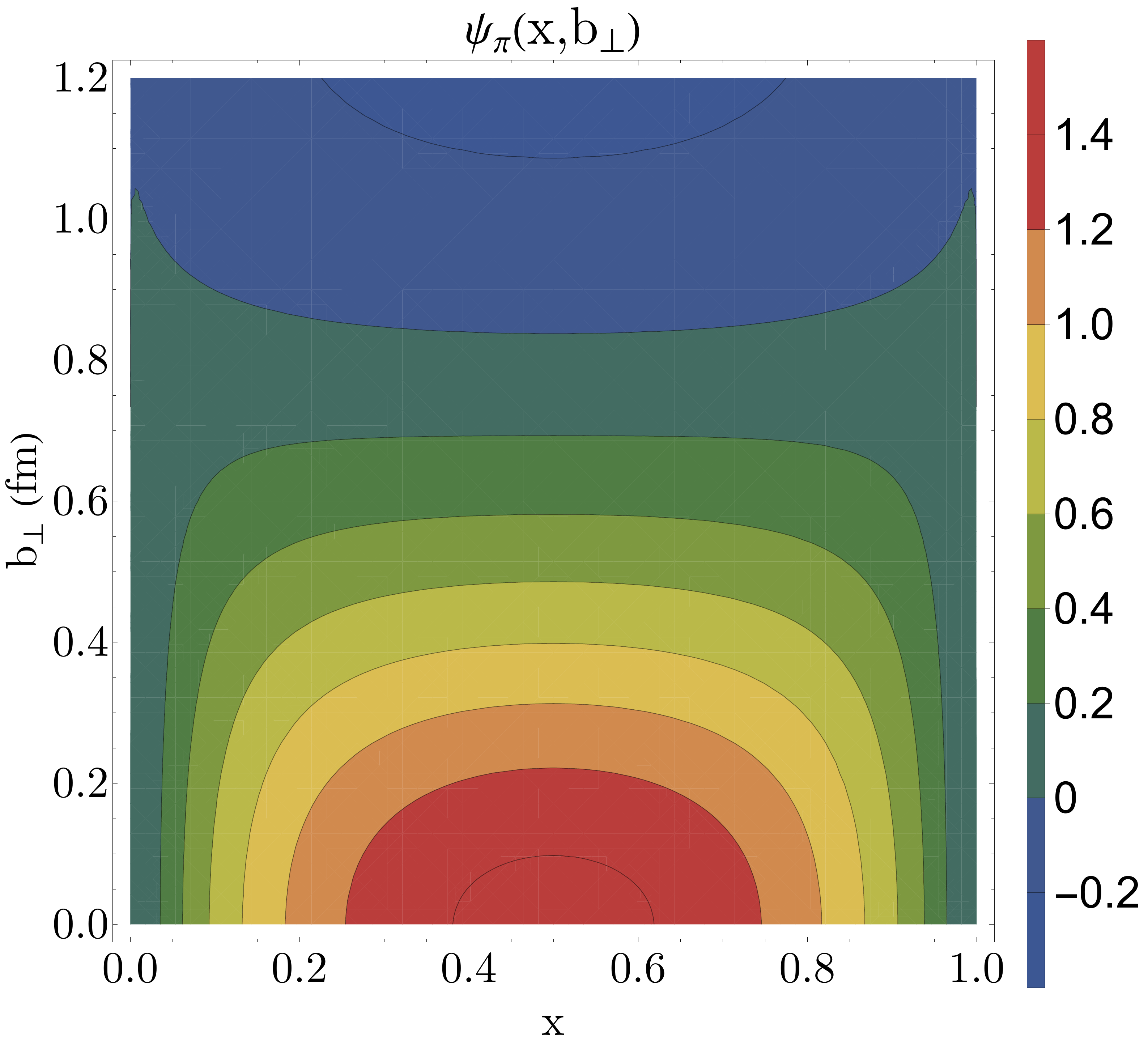}
\end{subfigure}%
\begin{subfigure}{0.5\textwidth}
  \centering
  \includegraphics[height=50mm]{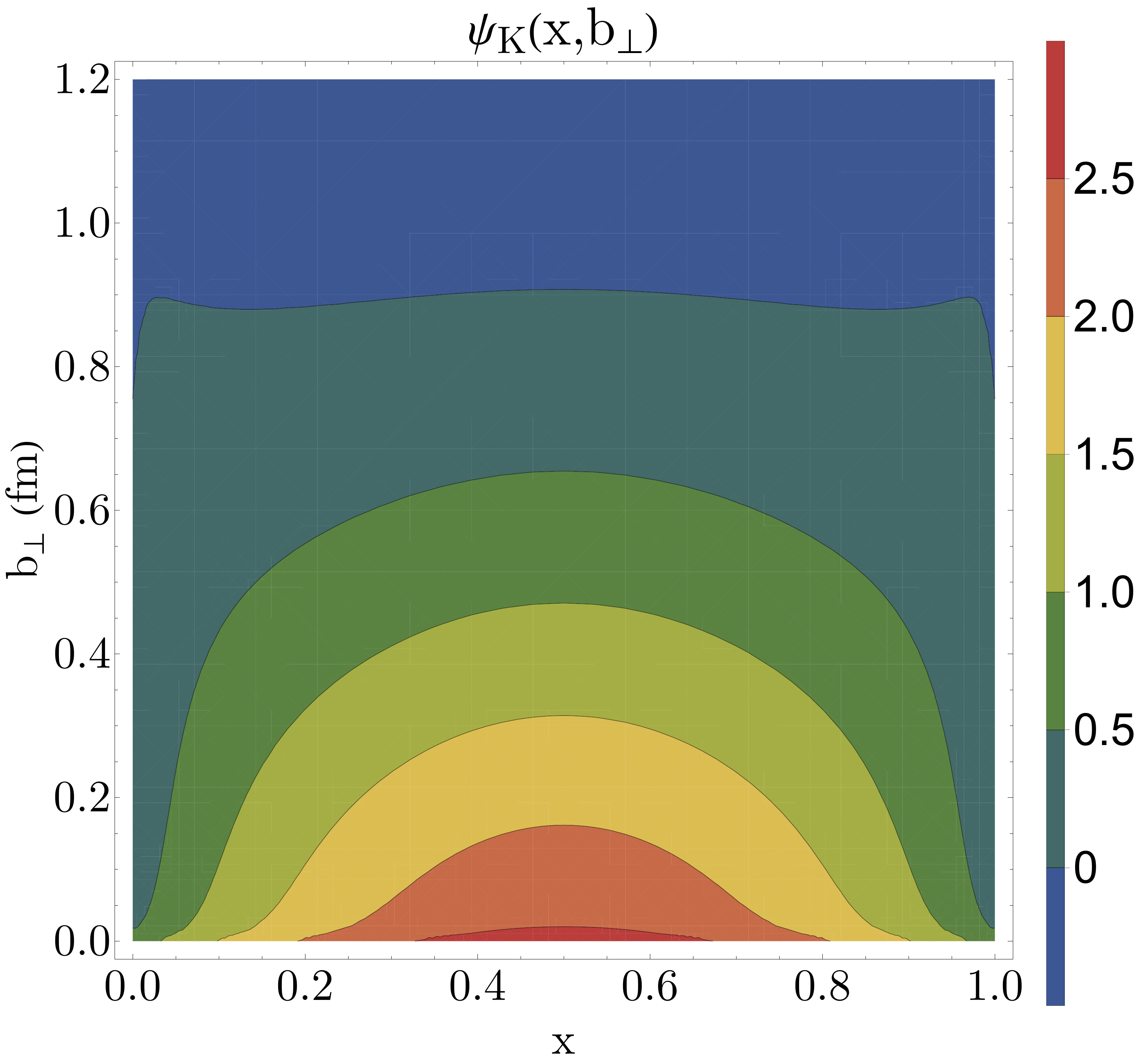}
\end{subfigure}

 \caption{Pion and Kaon transverse spatial distribution from the QCD instanton vacuum (\ref{GPSIZERO1}) with physical masses and at renormalization scale $Q_0=631\,\textrm{MeV}$.}
  \label{fig_tmdB}
\end{figure}

\begin{figure}[!htb]
\centering
\begin{subfigure}{0.5\textwidth}
  \centering
  \includegraphics[height=50mm]{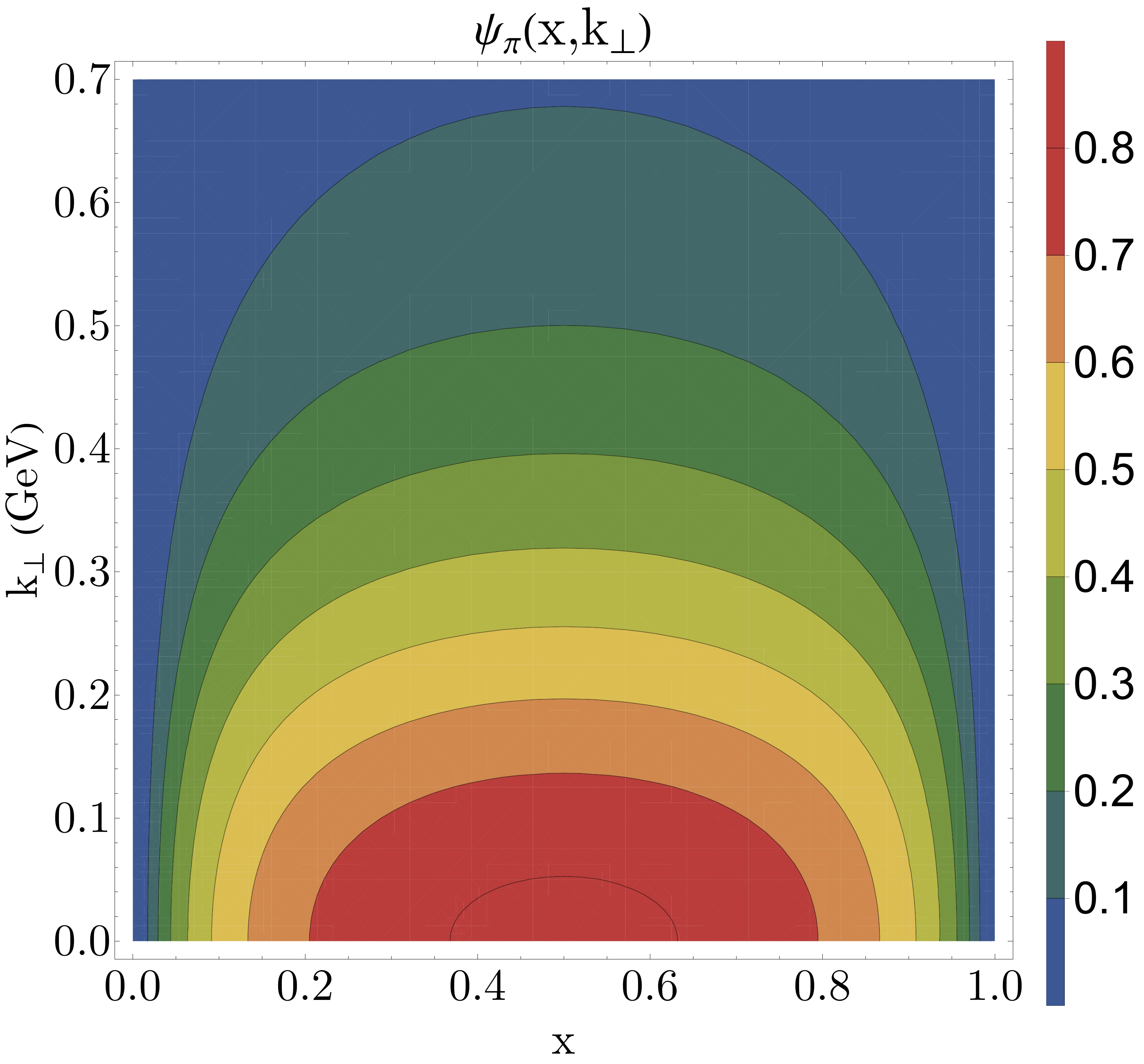}
\end{subfigure}%
\begin{subfigure}{0.5\textwidth}
  \centering
  \includegraphics[height=50mm]{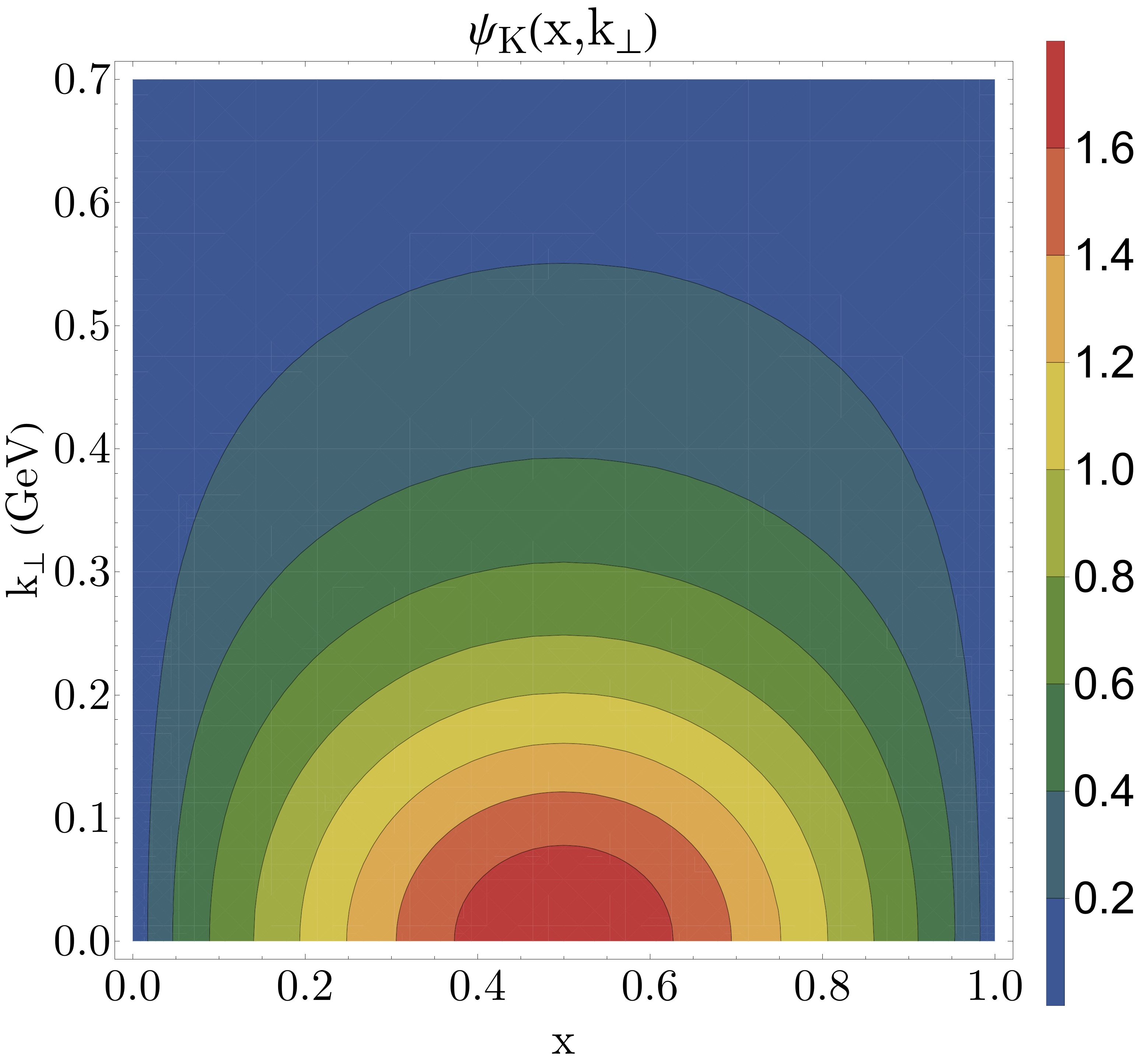}
\end{subfigure}

 \caption{Pion and Kaon transverse momentum distribution from the QCD instanton vacuum (\ref{GPSIZERO1}) with physical masses and at renormalization scale $Q_0=631\,\textrm{MeV}$.}
  \label{fig_tmdP}
\end{figure}

\newpage

\end{widetext}



\section{Conclusions}

We  revisited the  QCD  instanton vacuum in the context of an exact planar re-summation
of the n-point correlations that preserves both gauge and chiral symmetry in power counting 
using the root of the packing
fraction $\alpha\approx \sqrt\kappa$. We analysed  the  induced quark mass, effective pion
pseudo-scalar and pseudo-vector vertices at NLO with full conformity with the axial Ward identity
in the chiral limit. NNLO contributions are readily available but tedious.

We used this framework to derive the soft contributions to the pion  and kaon QPDA, QPDF and
QGPDF  following from the QCD  instanton vacuum. The results in LO show that these
pion quasi-distributions  receive contributions from both the zero modes and the non-zero modes,
but the latters drop out in the large momentum limit from the PDA, PDF  and TMD.
They are made explicit at LO or leading logarithm approximation.

The results we presented for the pion and kaon partonic distributions are all evaluated at the
low renormalization scale set by the inverse instanton size $1/\rho= 631$ MeV. A more compelling comparison with data at larger scales
require perturbative QCD evolution, assuming that factorization holds at this relatively low scale. Good agreements with the existing data
for the pion PDA was found for all-x, and the pion PDF  at moderate-x. 

The present analysis of the pion and kaon quasi-parton distributions
relies on a diagrammatic expansion and power counting in $\alpha$ to enforce  chiral and gauge
symmetry. It can be extended to all orders in $\alpha$ using  well tested numerical simulations for the QCD instanton vacuum,
that we will present elsewhere. In this respect, cooled lattice simulations of quasi-parton distributions which are expected
to be less noisy than the current simulations, would be welcome for comparison.
The present results can be extended to the  baryons away from the chiral limit. 

One of the chief proposal for the forthcoming EIC is the understanding of the origin of mass
and spin in most visible matter and its budgeting in terms of the fundamental constituents. The arguments we 
presented for the pion  and kaon,  show explicitly how most of their  composition is due to light quarks 
rescattering in a randomly distributed and non-perurbative sea of localized gluons in the form of instantons and anti-instantons.
The result is a running effective quark mass  that dictates how
the partons are distributed transversely in the light cone limit at the low renormalization scale. The magnitude of this effective mass
is a measure of the instanton-anti-instanton packing fraction in the QCD vacuum. It  is strongly dependent on the 
instanton size in the ultra-violet, and weakly dependent on the light current $u,d,s$ quark masses in the infrared. The collectivization
of the light quark zero modes when properly continued to the light cone through the large momentum limit, dominates the  light mesons leading twist contributions
 thanks to the diluteness of the QCD instanton vacuum.

Standard lore  says that in  light front  quantization (LFQ) the vacuuum is ${\it trivial}$~\cite{TRIVIAL}.  So how do we  reconcile 
this with the present arguments that show that the quasi-parton distributions  for the light mesons carry vacuum physics all
the way to the  infinite momentum limit? The answer lies in the neglected zero modes which when carefully treated in lower dimensions reproduce 
the chiral condensate in LFQ~\cite{LFQCHIRAL}. Recently, these zero modes were argued to pile up at zero x-parton~\cite{LFQJI},
much like a superfluid component in the otherwise normal fluid light cone wavefunction, 
and show up as  singular distributions in  higher twist observables as noted in~\cite{LFQBUR}. 
Recall that the chiral condensate observed here as a twist three operator, is a scalar in all frames, including the
light cone frame. It will be interesting to address the higher twist distributions in the present context.


\section{Acknowledgements}
We thank Edward Shuryak and Xiangdong Ji for discussions.
This work was supported by the U.S. Department of Energy under Contract No.
DE-FG-88ER40388, and by the Science and Technology Commission of Shanghai Municipality (Grant No.16DZ2260200).

\appendix

\section{Zero modes and non-zero mode quark propagator}

In singular gauge, the  instanton and anti-instanton quark zero modes in momentum space  are  locked in color-spin with a specific
chirality

\bea
\label{4X11}
&&\psi_{0I,\bar I}(p)=\sqrt{2}\varphi^\prime (p)\slashed{\hat p}\,\chi^{\pm}\nonumber\\
&&\varphi^\prime(p)=\pi\rho^2\bigg(I_0(z)K_0(z)-I_1(z)K_1(z)\bigg)^\prime_{z=p\rho/2}\nonumber\\
\eea
The prime is a z-derivative and  $I,K$ are modified Bessel functions.  The  corresponding zero mode projectors are

\be
\label{9X}
P_I(p)=2 \varphi^{\prime 2} \slashed{\hat p}\chi^+{\overline\chi}^-\slashed{\hat p}=\frac{\varphi^{\prime 2}(p)}{8p^2}\,\tau_\mu^-\tau_\nu^+\slashed{p}\gamma_\mu\gamma_\nu\slashed{p}\frac {1-\gamma_5}2
\nonumber\\
P_{\bar I}(p)=2\varphi^{\prime 2} \slashed{\hat p}\chi^-{\overline\chi}^+\slashed{\hat p}=\frac{\varphi^{\prime 2}(p)}{8p^2}\,\tau_\mu^+\tau_\nu^-\slashed{p}\gamma_\mu\gamma_\nu\slashed{p}\frac {1+\gamma_5}2\nonumber\\
\ee
with $\tau^\pm_\mu=(\vec \tau, \mp i)$. For comparison, note that the zero modes in regular gauge are simpler

\bea
\label{4X11X}
\psi_{0I,\bar I}(p)=\sqrt{2}\varphi^\prime (p)\chi^{\pm}\qquad \varphi^\prime (p)=4\pi\rho^2 \,e^{-p\rho}/(p\rho)\nonumber\\
\eea

The non-zero mode are more involved to construct, but a closed form for their propagator is known
 in singular gauge~\cite{BROWN}

\begin{widetext}
\be
\label{GIXY}
&&G_I(x,y)=\gamma_\mu D_\mu^x\Delta_{+}(x,y)\frac{1+\gamma_5}2\
+\Delta_{+}(x,y)\gamma_\mu D_\mu^y\frac{1-\gamma_5}2\nonumber\\
&&G_{\bar I}(x,y)=\gamma_\mu D_\mu^x\Delta_{-}(x,y)\frac{1-\gamma_5}2
+\Delta_{-}(x,y)\gamma_\mu D_\mu^y\frac{1+\gamma_5}2\nonumber\\
&&\Delta_\pm(x,y)=\frac 1{4\pi^2(x-y)^2}
 \bigg(1+\frac {\rho^2}{x^2}\bigg)^{-\frac 12}\bigg(1+\frac{\rho^2\tau^{\mp}_\mu \tau^\pm_\nu x_\mu y_\nu}{x^2y^2}\bigg)
\bigg(1+\frac {\rho^2}{y^2}\bigg)^{-\frac 12}
\ee
\end{widetext}
with the long derivative $D_\mu=\partial_\mu-iA_\mu$.
Both at short and large distances (\ref{GIXY}) reduce to the free propagator, while at intermediate distances it
is modified. More specifically,

\be
\label{APP}
G_{I}(x,y)\approx -\frac 1{2\pi^2}\frac{\gamma\cdot (x-y)}{(x-y)^2}-\frac 1{16\pi^2}
\frac{(x-y)_\mu\gamma_\nu\gamma_5}{(x-y)^2}\,\tilde F_{\mu\nu}\nonumber\\
\ee
with $\tilde{F}$ the dual of $F$.
All omitted terms in (\ref{APP}) are regular in the coincidental limit $x\rightarrow y$.

\begin{widetext}
\section{Pseudo-vector pion source and Axial Ward identity}

In this Appendix we detail the construction of the pseudo-vector pion source and show that it obeys
a pertinent axial Ward identity at LO. The re-summed planar approximation satisfies
the strictures of gauge and chiral symmetry.

\subsection{Axial-vector pion vertex at LO}

For the pion axial correlator we insert

\be
\label{18}
O^{\mu 5}(P,k)=\gamma^{\mu}\gamma^5+ \alpha F^{\mu 5}(P,k)+{\cal O}(\alpha^2)
\ee
in (\ref{13}), and use the LO contribution  for the quark propagator in (\ref{3}) and the NLO
contribution for the spin-valued self-energy(\ref{NLO3}). Power matching in $\alpha$ yields the spin-valued
integral equation

\be
\label{19}
&&F^{\mu 5}(P,k)=K_{\pi} F^{\mu 5}(P,k)\nonumber\\
&&+ \sum_{I, \bar I}\int \frac{d^4p}{(2\pi)^4}
\bigg( {\rm Tr}_C \Sigma_{I0}(k,p) \frac{i\sigma}{p^2}\gamma^{\mu}\gamma^5 S_0(p_-)\Sigma_{I0}(p_-,k_-)
+{\rm Tr}_C\Sigma_{I0}(k,p)  S_0(p)\gamma^{\mu}\gamma^5 \frac{i\sigma}{(p_{-})^2}\Sigma_{I0}(p_-,k_-)\nonumber \\
&&\qquad\qquad\qquad  +{\rm Tr}_C\Sigma_{I1}(k,p) S_0(p)\gamma^{\mu}\gamma^5 S_0(p_-)\Sigma_{I0}(p_-,k_-)
+{\rm Tr}_C\Sigma_{I0}(k,p) S_0(p)\gamma^{\mu}\gamma^5 S_0(p_-)\Sigma_{I1}(p_-,k_-)\bigg)\nonumber\\
\ee
Here $p_-=p-P$. The reduced kernel $K_{\pi}$ involves
only the zero modes and  satisfies

\be
\label{20}
K_{\pi} O= \sum_{I,\bar I}\int \frac{d^4p}{(2\pi)^4} {\rm Tr}_C\bigg( \Sigma_{I0}(k,p) S_0(p)O S_0(p_-)\Sigma_{I0}(p_-,k_-)\bigg)\nonumber\\
\ee
The $\beta_{00}$ contribution  in $\Sigma_{I1}$ in (\ref{NLO3})
does not contribute to this order, and the $(i\slashed{\partial}P_I \hat \sigma +i\hat \sigma P_{I}\slashed{\partial})$
contribution cancels exactly the first two terms in (\ref{19}). The final relation for $F^{\mu 5}$ simplifies

\be
\label{21}
&&F^{\mu 5}(P,k)=K_{\pi} F^{\mu 5}(P,k)\nonumber\\
&&+ \sum_{I+\bar I} \int \frac{d^4 p}{(2\pi)^4}
\bigg({\rm Tr}_C \bigg(\slashed{k}\tilde G_I(k,p)\gamma^{\mu}\gamma^5\psi_{0}(p_{-})\psi_{0}^{\dagger}(k_-)\frac{{\slashed{k}}_-}{i\sigma_{00}}\bigg)
+ {\rm Tr}_C \bigg(\frac{\slashed{k}}{i\sigma_{00}}\psi_{0}(k)\psi_{0}^{\dagger}(p)\gamma^{\mu}\gamma^5\tilde G^I(p_-,k_-)\slashed{k}_{-}\bigg)\bigg)\nonumber\\
\ee
\end{widetext}
Here

\be
\label{22}
\tilde G_I=(1-P_I\hat \sigma)G_I(1-\hat \sigma P_I)-S_0
\ee
is the projected and subtracted non-zero mode propagator which is UV finite.
The only non-vanishing contributions  to (\ref{21}) are

\be
\label{23}
-P_I \hat \sigma G_I-G_I\hat \sigma P_I +G_I-S_0
\ee
If  we aproximate $G_I \approx S_0$, then (\ref{21})  will reduce to the first two contributions in (\ref{19}) only.
This corresponds to expanding the propagator to first order while maintaining all  $\Sigma^\prime$s unchanged.  However, this
approximation upsets the axial Ward identity.

\begin{widetext}

\subsection{Axial Ward identity at LO}

In the chiral limit the pseudovector pion vertex  satistifies the exact Ward identity

\be
\label{24}
P_{\mu}O^{\mu 5}(k,P)=\gamma ^5S^{-1}(k-P)+S^{-1}(k)\gamma^5
\ee
to all orders in $\alpha$, which guarantees the transversality of the
 the axial-vector correlator

\be
\label{25}
P_{\nu}\left<F^{\mu 5}(-P)F^{\nu 5}(P)\right>=
\int \frac{d^4k}{(2\pi)^4}{\rm Tr}_C\bigg(\gamma^{\mu}\gamma^5 (S^{-1}(k)\gamma^5+\gamma^5 S^{-1}(k-P))\bigg)=0
\ee
since $S^{-1}(k)=\slashed{k}-i\sigma(k)$. The enforcement of the Ward identity and power counting guarantees
chiral and gauge symmetry. In particular, the extraction of the pion decay constant in power counting whether
from the pseudoscalar vertex or the pseudovector vertex is unique order by order in $\alpha$. This is not the case
in the partial resummations used in~~\cite{DP,MF}  where different values of $f_\pi$ were noted. Since the normalization
of the PDA and PDF involve $f_\pi$, the strict enforcement of the Ward identities is required.

(\ref{24}) fixes
uniquely the longitudinal part of the pseudovector pion vertex to all orders in $\alpha$

\be
\label{23X}
F_L^{\mu 5}(k,P)=-i\gamma_5(\sigma(k)+\sigma(k-P))\frac{P_\mu}{P^2}
\ee
In contrast, the transverse part is more involved, and can be only obtained through an expansion, with in LO

\be
\label{Z38}
F^{\mu 5}(k, P)&&=\lambda(P)\left<0|\delta F^{\mu 5}(k, P)|0\right>
\gamma_5 |k||k-P|\varphi^{\prime}(k)\varphi^{\prime}(k-P) +\delta F^{\mu 5}(k,P)
\ee
Here $\delta F^{\mu 5}$ refers to the inhomogenous contribution in  (\ref{21}) to order $\alpha$.
The pion pole resides in $\lambda (P)$ with  $\delta F^{\mu 5}$ regular at $P^2=0$.
Since  $\left<0|\delta F^{\mu 5}|0\right>$ is of the form $\sim k^\mu\gamma^5$ or $ P^\mu\gamma^5$,
it follows that the axial-axial vector correlation function  vanishes to order $\alpha$. We expect the axial-axial
vector correlator to be transverse and of order $f_\pi^2\sim \alpha^2$, as we now show.


We now proceed to show that our power counting enforces (\ref{25}) order by order.
For that, consider  the  contribution $-P^{I}\hat \sigma G^I-G^I\hat \sigma P^I$ in the inhomogeneous part of (\ref{21}),
   and contract it with $P_\mu$ . The result is

\be
\label{26}
-{\rm Tr}_C\slashed{p}\psi_{I0}(p)A_I^{\mu}(P)\psi_{I0}^{\dagger}(p-P)\frac{\slashed{p}-\slashed{P}}{i\sigma_{00}}
-{\rm Tr}_C\slashed{p}\psi_{I0}(p)B_I^{\mu}(P)\psi_{I0}^{\dagger}(p-P)\frac{\slashed{p}-\slashed{P}}{i\sigma_{00}}
\ee
where we have defined


\be
\label{27}
&&A_I^{\mu}=\int \frac{d^4 k}{(2\pi)^4}\beta^{\dagger}(p)\gamma^{\mu}\gamma^5 \psi_{I0}(k-P)\nonumber\\
&&B_{I}^{\mu}=\int \frac{d^4 k}{(2\pi)^4}\psi_{I0}^{\dagger}(p)\gamma^{\mu}\gamma^5 \beta(p-P)
\ee
with

\be
\label{28}
\beta (p)= \int \frac{d^4 k}{(2\pi)^4}G_I(p,k)\hat\sigma (k) \psi_{I0}(k)
\ee
or equivalently (x-space)

\be
\label{29}
A_{I}=\int d^4x \,\beta^{\dagger}(x)\gamma^{\mu}\gamma_5 \psi_{I0}(x)e^{i Px}\nonumber\\
B_{I}=\int d^4\, x \psi_{I0}^{\dagger}(x)\gamma^{\mu}\gamma_5 \beta(x) e^{iPx}
\ee
so that

\be
\label{30}
&&P_{\mu}B_{I}^{\mu}=\int d^4 x (-i\slashed{D_I}\psi_{0I})^{\dagger}\gamma_5 \beta(x)e^{iPx} -\int d^4 x\psi_{0I}^{\dagger}\gamma_5 i\slashed{D_I}\beta(x)e^{iPx}\nonumber\\
&&P_{\mu}A_{I}^{\mu}=\int d^4 x (-i\slashed{D_I}\beta)^{\dagger}\gamma_5 \psi_{0I}(x)e^{iPx} -\int d^4 x\beta^{\dagger}\gamma_5 i\slashed{D_I}\psi_{0I}(x)e^{iPx}
\ee
From (\ref{28}) it follows that  $\beta(x)=\int d^4d d^4 yG^{I}(x,z)\hat \sigma(z-y)\psi_0(y)$, so that  the action of $\slashed D_I$ on $\beta(x)$ is fixed by the zero mode only. Similarly for the contribution with $G^I$, which gives

\be
\label{31}
{\rm Tr}_C\slashed{p}C_I^{\mu}\psi_{0I}^{\dagger}(p-P)\frac{\slashed{p}-{\slashed P}}{i\sigma_{00}}
+{\rm Tr}_C\frac{\slashed{p}}{i\sigma_{00}}\psi_{0I}(p)D^{\mu}_I(\slashed{p}-{\slashed P})\nonumber\\
\ee
with

\be
\label{32}
&&C_I^{\mu}=\int d^4y \, G^I(p,y)\gamma^{\mu}\gamma^5 \psi_{0I}(y)e^{iPy}\nonumber\\
&&D_I^{\mu}=\int d^4y\,\psi_{0I}^{\dagger}\gamma^{\mu}\gamma^5 G^I(y,p-P)e^{iPy}
\ee
They  can be simplified using the same observations. Hence, after contracting with $P_\mu$ the results are

\be
\label{33}
&&P_{\mu}(A_{I}^{\mu}+B_I^{\mu})=
\int \frac{d^4p}{(2\pi)^4}\psi_{0I}^{\dagger}(p)\gamma^5 \psi_{0I}(p-P)
\bigg(2-\frac{\sigma_p+\sigma_{p-P}}{\sigma_{00}}\bigg)\nonumber\\
&&P_{\mu}C_I^{\mu}=-\gamma_5\psi_{0I}(p-P)+\psi_{0I}(p)\int\frac{d^4k}{(2\pi)^4}\psi_{0I}^{\dagger}(k)\gamma^5 \psi_{0I}(k-P)\nonumber\\
&&P_{\mu}D_I^{\mu}=-\psi^{\dagger}_{0I}(p)\gamma_5
+\int\frac{d^4k}{(2\pi)^4}\psi_{0I}^{\dagger}(k)\gamma^5 \psi_{0I}(k-P)\psi_{0I}^{\dagger}(p-P)
\ee
Finaly, the contribution with $1/i\slashed{\partial}$ can be direcltly calculated and gives after contracting with $P_{\mu}$

\be
\label{34}
{\rm Tr}_C\slashed{P}\gamma^5\psi_{0I}(p-P)\psi_{0I}^{\dagger}(p-P)\frac{\slashed{p}-{\slashed P}}{i\sigma_{00}}
-{\rm Tr}_C\frac{\slashed{p}}{i\sigma_00}\psi_0(p)\psi_{0I}^{\dagger}(p)\gamma^5\slashed{P}
\ee
While combining the above results, we note that the second term in $C$ and $D$ cancel with the 2 in
the bracket  $(2-{\sigma}/{\sigma_{00}})$ for $A$ and $B$, and the
contributions $\gamma^5 \psi_0$  and  $\psi^{\dagger}_0\gamma^5$ in $C$ and $D$ respectively,
 combine with the  contribution  ${1}/{i\slashed{\partial}}$ to give $\slashed{p}-{\slashed P}$ or $\slashed{p}$ respectively. The final result after contracting with $P_{\mu}$ is

\be
\label{35}
&&\sum_ I {\rm Tr}_C \bigg(\slashed{p} \psi_{0I}(p)\bigg(
 \int \frac{d^4 k}{(2\pi)^4}\psi^{\dagger}_{0I}(k)\frac{\sigma_k+\sigma_{k-P}}{\sigma_{00}}\gamma_5
\,\psi_{0I}(k-P)\bigg)\psi^{\dagger}_{0I}(p-P)\frac{\slashed{p}-{\slashed P}}{i\sigma_{00}}\bigg)\nonumber \\
+&&\sum_I {\rm Tr}_I\bigg(\gamma_5(\slashed{p-P})\frac{\psi_{0I}(p-P)\psi_{0I}^{\dagger}(p-P)}{i\sigma_{00}}(\slashed{p}-{\slashed P})\bigg)
+\sum_I {\rm Tr}_C\bigg(\gamma_5\slashed{p}\frac{\psi_{0I}(p)\psi_{0I}^{\dagger}(p)}{i\sigma_{00}}\slashed{p}\bigg)
\ee
Using the definition of $K_{\pi}$ in (\ref{20}) and the gap equation for $\sigma$,  (\ref{35}) can be written as

\be
\label{36}
(1-K_{\pi})(-i\gamma^5 \sigma(k)-i\gamma^5 \sigma(k-P))
\ee
which is the action of $P_\mu$ on the inhomogeneous part of (\ref{21}), or

\be
\label{37}
(1-K_{\pi})(P_{\mu}F^{\mu 5}(k,P))=
(1-K_{\pi})(-i\gamma^5 \sigma(k)-i\gamma^5 \sigma(k-P))
\ee
It follows that

\be
\label{38}
P_{\mu}F^{\mu 5}(k,P)=-i\gamma^5 \sigma(k)-i\gamma^5 \sigma(k-P)
\ee
which is the axial Ward identity expanded to first order in $\alpha$.
This concludes our proof that (\ref{18})
and the corresponding 2-point correlation function satisfies  the axial Ward identity at LO in  $\alpha$.

\section{Gauge link}

We can show that to the same order $\alpha^0$ the only contribution of the
gauge link in (\ref{X1}) follows from (\ref{X7X}) with the substitution

\be
\label{X8X}
\int \frac{d^3 p}{(2\pi)^3}\psi^{\dagger }_{0I}(p^-)\gamma^z \gamma^5 \delta G_{I}(p,k)\rightarrow
\int d^4 y dz e^{-iPy+iP_z(x-\frac 12)z}
 \psi^{\dagger }_{0I}\bigg(y+\frac{z}{2}\bigg)\gamma^z\gamma^5\left[y+\frac{z}{2},y-\frac{z}{2}\right]_I\delta G_{I}\bigg(y-\frac{z}{2},k\bigg)
\ee
for the first term, and similarly for the second term. The gauge link involves the z-propagation of a quark
in a single instanton,

\be
\label{X9X}
\bigg[y+\frac z2,y-\frac z2\bigg]_I=\left<y+\frac z2 \left|\frac 1{i\partial_z-A_{Iz}}\right|y-\frac z2\right>
\ee
and restores explicit gauge invariance in (\ref{X7X}) to order $\alpha^0$. For instance, in the regular gauge
with  $A_{M}=-\bar \sigma_{MN}x_N\frac{1}{x^2+\rho^2}$, the gauge  link simplifies

\be
\left[y+\frac{z}{2},y-\frac{z}{2}\right]_{I,\bar I}
=\cos F(r_3,y_z,z) \pm i\sigma \cdot \hat r_3 \sin F(r_3,y_z,z)
\ee
with

\be
F(r_3,y_z,z)= \int_{-1}^{1}d\tau\, \left[\frac{r_3 \frac{z}{2}}{r_3^2+\rho^2+(y_z+\frac{\tau z}{2})^2}\right]
=\frac{r_3}{\sqrt{r_3^2+\rho^2}}\left[{\rm arctan}\bigg(\frac{y_z+\frac{z}{2}}{(r_3^2+\rho^2)^{\frac{1}{2}}}\bigg)-{\rm arctan}\bigg(\frac{y_z-\frac{z}{2}}{(r_3^2+\rho^2)^{\frac{1}{2}}}\bigg)\right]\nonumber \\
\ee

\section{Generalized QPDA}

The pion QPDA (\ref{7}) in the random instanton vacuum is part  of a larger class of quasi-distributions.
For instance, the LO contribution (\ref{PHI02}) can be recast in the general form

\be
\label{8X}
\tilde{\phi}^0_\pi(x,n,P)=-\frac{4iN_c}{f_\pi^2}\int \frac{d^4k}{(2\pi)^4}\delta\left(n\cdot k-\left(x-\frac{1}{2}\right)n\cdot P\right)\,
(M(y_1)M(y_2))^{\frac 12}\,
 \frac{(n\cdot p_1M(y_2)+n\cdot p_2 M(y_1)) }{y_1^2y_2^2}
\ee
with  $n$ an arbitrary 4-vector,   using Minkowski signature and the causal assignment for the poles.
Lorentz and $^{\prime\prime}$scale$^{\prime\prime}$ invariance imply

\be
\label{9X}
\tilde{\phi}^0_\pi(x,n,P)\equiv \phi_\pi\bigg(x,P^2,\frac{n^2}{(n\cdot P)^2}\bigg)
\ee
For time-like $n=n_-$ and space-like $n=n_z$ we have respectively the PDA and QPDA, i.e.

\be
\label{10X}
\phi^0_\pi(x)=\tilde{\phi}^0_\pi(x,m_{\pi}^2,0)\qquad\qquad
\tilde{\phi}^0_\pi(x,P_z)=\tilde{\phi}^0_\pi\bigg(x,m_{\pi}^2,\frac{1}{P_z^2}\bigg)
\ee
For large $P_z$, the pion QPDA reduces to the PDA in the random instanton vacuum.


\section{Zero and non-zero modes on-shell}

The zero and non-zero modes entering our analysis of the PDA and PDF simplify when they are put on
mass shell which is the leading contribution for the quasi-distributions in the large $P_z$ limit. For the
zero modes, the mass-shell reduction yields constant Weyl spinors. From (\ref{4X11}) we have for the zero modes

\be
\label{RULE1}
\slashed p\psi_{0I,0\bar I}(p)\rightarrow -\sqrt{2}\pi\rho\chi^\pm
\ee
with $p^2=0$. For the non-zero modes we have 

\be
\label{RULE2}
&&\int d^4x \,e^{-iq\cdot x}\psi_{0I}^+(x)\bar{\sigma}_z\delta G_I(x,p)\,i\overline{p}\rightarrow {\mathbb F}(q,p)\nonumber\\
&&\int d^4x \,e^{-iq\cdot x}\,i{p}\,{\delta \bar G_I}(p,x) {\sigma}_z\psi_{0I}(x)\rightarrow \overline{\mathbb F}(q,p)
\ee
with $\delta\bar G_I$ following from $\delta G_I$ by barring the Weyl contributions. 
A tedious derivation following the arguments presented in~\cite{RINGWALD} gives
for the mixed and subtracted instanton non-zero mode contributions the results in (\ref{ONSHELL1}).

\section{Pion QGPDF at LO}

The pion quasi-generalized distribution function (QGPDF) can also be extracted from the equal-time correlator
following (\ref{1}) as suggested in~\cite{JI1}, with formally

\be
\label{Z1}
\tilde \Psi_\pi(x, q, P)=\int \frac{dz}{2\pi}\,e^{-i(x-\bar x)zp_z}\,
 \left<\pi(P+q)\right|\overline{\Psi}(z_-)\gamma^z\,[z_-,z_+]\,
\Psi(z_+)\left|\pi(P)\right>
\ee
In the random instanton vacuuum (\ref{Z1}) follows from the same reduction rules as those for the QDA and QPDA detailed above.
Both the zero modes and non-zero modes contribute to order $\alpha^0$ to LO, but the dominant contribution stems from
the zero modes in the large momentum limit as we noted earlier. The LO  result for the QGPDF after spin-color contractions 
and the free approximation for the non-zero modes $\delta G_I\approx S_0$, is

\be
{\tilde\Psi}^0_\pi(x,q,P)\approx \frac{4iN_c}{f_\pi^2}\int \frac{dk_4d^2k_\perp}{(2\pi)^4}\,\left(M(k_1)M^2(k_2)M(k_3)\right)^{\frac 12}\,
\bigg(\frac{k_1^z+k_3^z}{2k_1^2k_3^2}+\frac{k_1^z}{2k_1^2k_2^2}+\frac{k_3^z}{2k_2^2k_3^2}-\frac{k_2^z}{k_2^2k_3^2}-\frac{k_2^zk_1\cdot q}{k_1^2k_2^2k_3^2}\bigg)+{\rm cross}\nonumber\\
\label{Z2}
\ee
with $k_\perp\geq 0$ subsumed, $k_z= xP_z$ and

\be
\label{Z3}
k_1=k-\frac{q}{2}\qquad
k_2=-P-\frac{q}{2}+k\qquad
k_3=k+\frac{q}{2}
\ee
The cross terms have the same structure but with the substitution $k_2\rightarrow k^{\prime}_2=P+k+\frac{q}{2}$.
The GPDF follows from (\ref{Z2}) in the large $P_z$  limit. It will  be analyzed elsewhere.

\end{widetext}

 \vfil

\end{document}